\documentclass[lettersize,journal]{IEEEtran}

\usepackage[colorlinks,linkcolor=blue,anchorcolor=blue,citecolor=blue]{hyperref}
\usepackage[nocompress]{cite}
\usepackage{algorithm, algorithmic}
\usepackage{amsmath,amssymb,amsfonts,mathrsfs}
\usepackage{amsthm}
\usepackage{array}
\usepackage{balance}
\usepackage{booktabs}
\usepackage{graphicx}
\usepackage{makecell}
\usepackage{siunitx}
\usepackage{stfloats}
\usepackage{subfigure,graphicx}
\usepackage{textcomp}
\usepackage{url}
\usepackage{verbatim}
\usepackage{xcolor}

\def\BibTeX{{\rm B\kern-.05em{\sc i\kern-.025em b}\kern-.08em
    T\kern-.1667em\lower.7ex\hbox{E}\kern-.125emX}}

\newcommand{\Tr}{{\rm Tr}}

\newcommand{\by}{\mathbf{y}}

\newcommand{\bV}{\mathbf{V}}

\newcommand{\bX}{\mathbf{X}}

\theoremstyle{definition}
\newtheorem{remark}{Remark}
\newtheorem{Proposition}{Proposition}
\newtheorem{theorem}{Theorem}
\newtheorem{lemma}{Lemma}
\begin{document}
\title{Joint Beamforming Design and Stream Allocation for Non-Coherent Joint Transmission in Cell-Free MIMO Networks}
\author{Xi Wang, Xiaotong Zhao, Juncheng Wang,  You Li, and
Qingjiang Shi
\thanks{Xi Wang and Xiaotong Zhao are with the School of Software Engineering,
Tongji University, Shanghai 201804, China (e-mail: \{wangxi\_w,xiaotongzhao\}@tongji.edu.cn).
Juncheng Wang is with the Department of Computer Science, Hong Kong Baptist University, Hong Kong, China (e-mail: jcwang@comp.hkbu.edu.hk).
You Li is with Huawei Technologies Co. Ltd, Chengdu 611730, China (e-mail: liyou1992@163.com).
Qingjiang Shi is with the School of Software Engineering, Tongji University,
Shanghai 201804, China, and also with the Shenzhen Research Institute of Big
Data, Shenzhen 518172, China (e-mail: shiqj@tongji.edu.cn).}
}

\markboth{Journal of \LaTeX\ Class Files,~Vol.~18, No.~9, September~2020}%
{How to Use the IEEEtran \LaTeX \ Templates}

\maketitle

\begin{abstract}
We consider joint beamforming and stream allocation to maximize the weighted sum rate (WSR) for non-coherent joint transmission (NCJT) in user-centric cell-free MIMO networks, where distributed access points (APs) are organized in clusters to transmit different signals to serve each user equipment (UE). 
We for the first time consider the common limits of maximum number of receive streams at UEs in practical networks, and formulate a joint beamforming and transmit stream allocation problem for WSR maximization under per-AP transmit power constraints. 
Since the integer number of transmit streams determines the dimension of the beamformer, the joint optimization problem is mixed-integer and nonconvex with coupled decision variables that is inherently NP-hard. 
In this paper, we first propose a distributed low-interaction reduced weighted minimum mean square error (RWMMSE) beamforming algorithm for WSR maximization with fixed streams.
Our proposed RWMMSE algorithm requires significantly less interaction across the network and has the current lowest computational complexity that scales linearly with the number of transmit antennas, without any compromise on WSR.
We draw insights on the joint beamforming and stream allocation problem to decouple the decision variables and relax the mixed-integer constraints. We then propose a joint beamforming and linear stream allocation algorithm, termed as RWMMSE-LSA, which yields closed-form updates with linear stream allocation complexity and is guaranteed to converge to the stationary points of the original joint optimization problem.
Simulation results demonstrate substantial performance gain of our proposed algorithms over the current best alternatives in both WSR performance and convergence time.
\end{abstract}


\section{Introduction}
\IEEEPARstart{T}{he} 
exponential increase in the number of dense and heterogeneous terminals and their diverse requirements present major challenges to modern wireless communication networks, such as increased inter-cell interference and attenuated cell-edge rate. 
Due to the co-located antennas architecture of traditional cellular paradigm,  cell-edge user equipments (UEs) are inevitably far away from the base stations (BSs), and thus suffer from edge rate attenuation. 
The paradigm of cell-free  multiple-input multiple-output (MIMO) was pioneered in~\cite{yang2013capacity} to provide uniform service to all the UEs. 
In cell-free MIMO networks, geographically distributed access points (APs) are coordinated by one or multiple central units (CUs) to jointly serve the UEs.
In such networks, all the UEs are ensured to be located at the effective center of their serving APs, and cell edges no longer exist.
Therefore, the inherent large data rate variation and inter-cell interference are mitigated\cite{ammar2021downlink}. 
Moreover, the proximity of densely deployed APs to their serving UEs in cell-free MIMO networks brings enhanced energy efficiency\cite{hoydis2011green}, low communication latency, and augmented service reliability \cite{zhang2019cell,papazafeiropoulos2021towards}. 
These exceptional benefits of cell-free MIMO networks position it as a promising paradigm for future wireless communication systems.

To provide high quality service and effectively manage interference, APs in cell-free MIMO networks need to perform joint transmission (JT) to enhance constructive signals and suppress destructive signals at the UEs\cite{6146494,6825087}. 
 There are two JT approaches for cell-free MIMO networks: coherent joint transmission (CJT) and non-coherent joint transmission (NCJT).
 In CJT, all the APs cooperate as a virtual MIMO system to transmit the same signals to their serving UEs. Therefore, CJT requires strict synchronization across the network\cite{8482453}. 
However, in practical cell-free networks, APs are coordinated by multiple CUs instead of a single CU to avoid single-point failure \cite{li2015joint,bjornson2019making}.
It is difficult to deploy multiple CUs in CJT due to its strict synchronization requirement across the network, especially when the serving APs of a UE are controlled by different CUs \cite{ammar2017dynamic}. 
NCJT differs from CJT in that the signals transmitted to a UE from its serving APs are different. 
Since each UE can decode the received signals independently, strict synchronization between CUs and APs is no longer required for NCJT \cite{8482453,van2016joint}. 

In order to perform JT, the APs and the CUs need to exchange channel state information (CSI) and cooperative beamforming matrices via fronthaul communication links\cite{demir2021foundations}.
However, both the CSI and the beamformer is of high dimension, imposing a large burden on the fronthaul.
In practical communication networks, wireless fronthaul is widely deployed due to its high deployment flexibility and low installation cost. 
Even at the millimeter wave frequency, the available fronthaul capacity is limited, resulting in a bottleneck on efficient JT in cell-free MIMO networks\cite{8482453}.
It is therefore of critical importance to take into account the interaction between the APs and CUs in cell-free MIMO networks.
\subsection{Motivations and Challenges}
Most existing algorithms on beamforming design for NCJT are centralized \cite{9556147,7497508,6632074,vu2020noncoherent,8482453,ammar2021downlink}, which suffer from high interaction  (requiring raw CSI exchange across the network) and high computational complexity in general~\cite{bjornson2019making}.
Furthermore, they did not consider the possibility of maximizing the weighted sum rate (WSR) through data stream allocation among the UEs. Since the number of received data streams at each UE is limited by its receive antennas\cite{epstein2009scalable},
while there are fluent APs possibly equipped with a large number of transmit antennas  around each UE in cell-free MIMO systems, there is a plenty of room to further improve the WSR by properly allocating data streams from the APs to each UE. 
However, \emph{there is a scarcity of existing literature on how to jointly allocate data streams and design cooperative beamformer for NCJT in cell-free MIMO systems.}

Due to the above discrepancies, in this work, we consider joint beamforming design and stream allocation to maximize the WSR of cell-free MIMO networks with NCJT, under individual transmit power constraints at the APs. We aim at developing low-interaction and low-complexity distributed joint beamforming and stream allocation algorithms. To achieve this goal, we must address several challenges:
1) Since the signal-to-interference-plus-noise ratio (SINR) of each UE is coupled among its serving APs, the CUs intrinsically require global CSI data to effectively mitigate inter-UE interference. 
However, directly exchanging of raw CSI data (the high-dimensional channel matrices) between the APs and the CUs is practically prohibited due to the limited fronthaul capacity in practical cell-free MIMO systems, while communicating partial CSI generally degrades the system performance \cite{ammar2021user}.
It is therefore challenging to reduce the interaction between the APs and CUs without sacrificing the WSR.
2) Beamforming design and stream allocation are intrinsically coupled and hard to be jointly optimized, since the dimension of the beamformer is determined by the number of data streams. 
Even with fixed transmit streams, the pure beamforming optimization problem remains NP-hard. 
3) Due to the unique sum-of-quadratic form of the signal covariance matrix in the WSR expression for NCJT, whether the efficient weighted minimum mean square error (WMMSE) approach for conventional CJT can be applied to reduce the computational complexity for beamforming design in NCJT remains an open problem \cite{8482453,vu2020noncoherent}.



\subsection{Contributions}
Different from existing centralized beamforming algorithms for NCJT that are of high interaction (requiring raw CSI exchange between the APs and the CUs) and high complexity, we propose a distributed low-interaction and low-complexity beamforming algorithm using the WMMSE techniques. Furthermore, we \emph{for the first time} study joint beamforming and stream allocation for user-centric cell-free MIMO networks with NCJT. Specifically, the main 
contributions of this paper are as follows:
\begin{enumerate}
\item \textbf{An Answer to Whether the WMMSE Approach is Applicable to NCJT:} 
Prior works have assumed that the WMMSE approach is not applicable to NCJT\cite{8482453,vu2020noncoherent}.
Based on unique observations on the structure of the WSR expression for NCJT, 
we equivalently transform the original WSR maximization problem into a standard WMMSE form, showing the applicability of the WMMSE approach to NCJT.
We then propose a centralized WMMSE based beamforming algorithm that has the current lowest computational complexity.
\item \textbf{A Distributed Low-interaction and Low-Complexity Beamforming Algorithm:}
We draw some unique observations on the beamformer structure for WSR maximization in cell-free networks MIMO with NCJT.
Based on these observations, we propose a low-interaction reduced WMMSE (RWMMSE) algorithm, which converges smoothly and achieves the same WSR as the centralized WMMSE method. 
Without sacrificing any WSR performance, our proposed RWMMSE algorithm yields much lower interaction (independent of the number of transmit antennas) than the centralized WMMSE algorithm.
Furthermore, our proposed RWWMSE algorithm achieves a computational complexity that scales linearly with $M$, which is much lower than the current lowest 
$\mathcal{O}\left(M^3\right)$ complexity in  \cite{ammar2021downlink}, where $M$ is the number of transmit antennas.
\item \textbf{A Joint Beamforming and Linear Stream Allocation Algorithm:} 
We for the first time consider the individual maximum number of receive streams constraints in NCJT, and formulate a joint beamforming and stream allocation problem to maximize the WSR of cell-free MIMO networks. 
Note that the beamforming and stream variables are coupled, in the sense that the beamforming dimension is determined by the number of streams. 
Moreover, the joint optimization problem is mixed-integer and nonconvex. 
We first introduce auxiliary stream indicator matrices to decouple the beamforming and stream variables.
We then utilize the unique quadratic-linear property of the stream indicator matrices to transform the decoupled quadratic 0-1 integer stream allocation problem into an equivalent linear form,  and relax it to a continuous optimization problem. 
We propose a joint beamforming and linear stream allocation algorithm, termed as RWMMSE-LSA, which consists of closed-form updates with linear stream allocation complexity and is shown to converge to the stationary points of the original joint optimization problem.
%
%
\item Our simulation results demonstrate that the proposed algorithms converge fast without any WSR sacrifice, while substantially reducing the computation time compared to the current best alternatives. 
We further compare the WSR between CJT and NCJT approaches, showing a balance between synchronization cost and WSR performance provided by NCJT. In addition, our proposed joint beamforming and stream allocation algorithm outperforms the pure beamforming algorithms.
\end{enumerate}


\emph{Notations:} 
The notation ${\bf{A}} \succ {\bf{0}}$ indicates positive definite. 
The notation $\operatorname{blkdiag}(\mathbf{A}_1,\dots,\mathbf{A}_n) $ denotes a block diagonal matrix with matrices $\mathbf{A}_1,\dots,\mathbf{A}_n$. 
The column space of ${\bf{A}}$ is the span of its column vectors.
The null space of ${\bf{A}}$ is the linear subspace of the domain of the mapping to the zero vector.
Orthogonal complement of the column space of ${\bf{A}}$ is defined by $\prod_{{\bf{A}}}^{\bot} \triangleq \mathbf{I}-\mathbf{A}\left(\mathbf{A}^H \mathbf{A} \right)^{-1}\mathbf{A}^H $.
The binary and complex space are denoted as $\mathbb{B}$ and $\mathbb{C}$.

\section{Related Work}\label{sec_related_work}
\subsection{Beamforming Algorithms}
\subsubsection{Centralized Beamforming Algorithms}
Most existing beamforming algorithms for NCJT are centralized, which demand massive interaction between the CUs and the APs and suffer from high computational cost\cite{9556147,6632074,vu2020noncoherent,8482453,ammar2021downlink}.
For dense small cell networks, a semi-definite relaxation (SDR) based algorithm was proposed to minimize power consumption under transmit rate constraints\cite{6632074}. Although the SDR based algorithm converges in polynomial time, it is not applicable to the WSR maximization problem. 
The SCA method was used in \cite{8482453} to relax the nonconvex WSR maximization problem to a second order cone programming (SOCP) problem.
Similar scheme was adopted in \cite{9556147} for non-orthogonal multiple access systems.
The SOCP problem in \cite{8482453} is solved via CVX \cite{grant2014cvx}, which involves the interior point method that causes high computational complexity.

The authors in \cite{ammar2021downlink} utilized the fractional programming (FP) approach and the block coordinate descent (BCD) method to optimize the beamformer for WSR maximization under the special case that each UE is equipped with a single receive antenna. 
The FP based algorithm in \cite{ammar2021downlink} has much lower complexity than the SCA based algorithms, but still suffers from a $\mathcal{O}\left(M^3\right)$ computational complexity
 that is still high especially when massive antennas are deployed.
Moreover, all of the aforementioned beamforming algorithms require the interaction of channel and beamforming matrices, both of which are related to the number of transmit antennas, which puts a significant burden on the fronthaul links.
 \subsubsection{Low-Interaction Beamforming Algorithms}
Great efforts have been made to develop low-interaction beamforming algorithms especially for CJT\cite{ammar2021user}.
Beamforming appraoches such as local minimum mean square error (MMSE) combining \cite{bjornson2019making}, weighted MMSE (WMMSE) \cite{ammar2021downlink,5756489}, local partial zero forcing, and local protective partial zero forcing \cite{alonzo2018energy},
reduce the interaction by limiting the CSI sharing.

The unique WSR structure (a summation term resulting from different data streams in the SINR numerator) of NCJT makes it challenging to directly extend the above-mentioned low-interaction beamforming algorithms to NCJT.
Authors in \cite{8482453} and \cite{vu2020noncoherent} imply that the widely deployed WMMSE approach is not applicable to NCJT, whereas we suggest otherwise in this paper with a unique but equivalent WMMSE reformulation of the original WSR maximization problem. 
For NCJT, \cite{vu2020noncoherent} used the inner approximation (InAp) method to relax the nonconvex WSR maximization problem to a convex approximation subproblem, which is then solved by the alternating direction method of multipliers (ADMM) method. 
The algorithm presented in \cite{vu2020noncoherent} only requires scalar interactions across the network. 
However, the scalar interaction is required for each inner ADMM iteration, and the two-layer iterative approach adopted in the algorithm results in unavoidably high complexity.
Furthermore, \cite{vu2020noncoherent} focuses on the single receive antenna case, and the interaction is no longer scalar for the general multiple receive antenna cases.
%
\subsection{Stream Allocation and User Scheduling}
As stated in \cite{epstein2009scalable}, \emph{the number of usable spatial streams should be less than both of the number of transmit and receive antennas}.
However, stream allocation is overlooked in prior works on cell-free networks, and the number of transmit streams is pre-defined.
Besides, distant APs occupy power and bandwidth but contribute little receive power due to pathloss in cell-free networks \cite{demir2021foundations,zhou2003distributed}.
Therefore, UEs are not necessary served by all the APs but only nearby APs, which is generally referred to as user-centric cell-free networks \cite{bjornson2019making,bjornson2020scalable}.

Existing works on user-centric cell-free networks define the serving cluster by serving distance \cite{ammar2019power} or follow the dynamic cooperation clustering framework\cite{ammar2021distributed,vu2020noncoherent,8482453}. 
The authors in \cite{ammar2021distributed} treated the mixed-integer clustering problem as an agent-task assignment problem, and used the Hungarian algorithm to solve it in polynomial time.
In \cite{vu2020noncoherent}, a branch reduce-and-bound (BRnB) framework was developed to find the global optimal serving cluster for WSR maximization but with exponential complexity.
In \cite{8482453}, a joint beamforming and user scheduling problem was formulated under limits on the maximum number of UEs served by each AP. The mixed-integer problem was transformed to $\ell_{0}$-norm and then approximated using weighted $\ell_{1}$-norm. 
Since the algorithm in~\cite{8482453} involves in two-layer iterations, the authors merge the two-layer iteration into a one-layer iteration to reduce the computational complexity.
However, the resulting one-layer iteration algorithm cannot converge strictly. 

%
%
%
%
\section{System Model and Problem Formulation} \label{sec_sys}
\begin{figure}[!t]
\centering
\includegraphics[width=0.9\columnwidth]{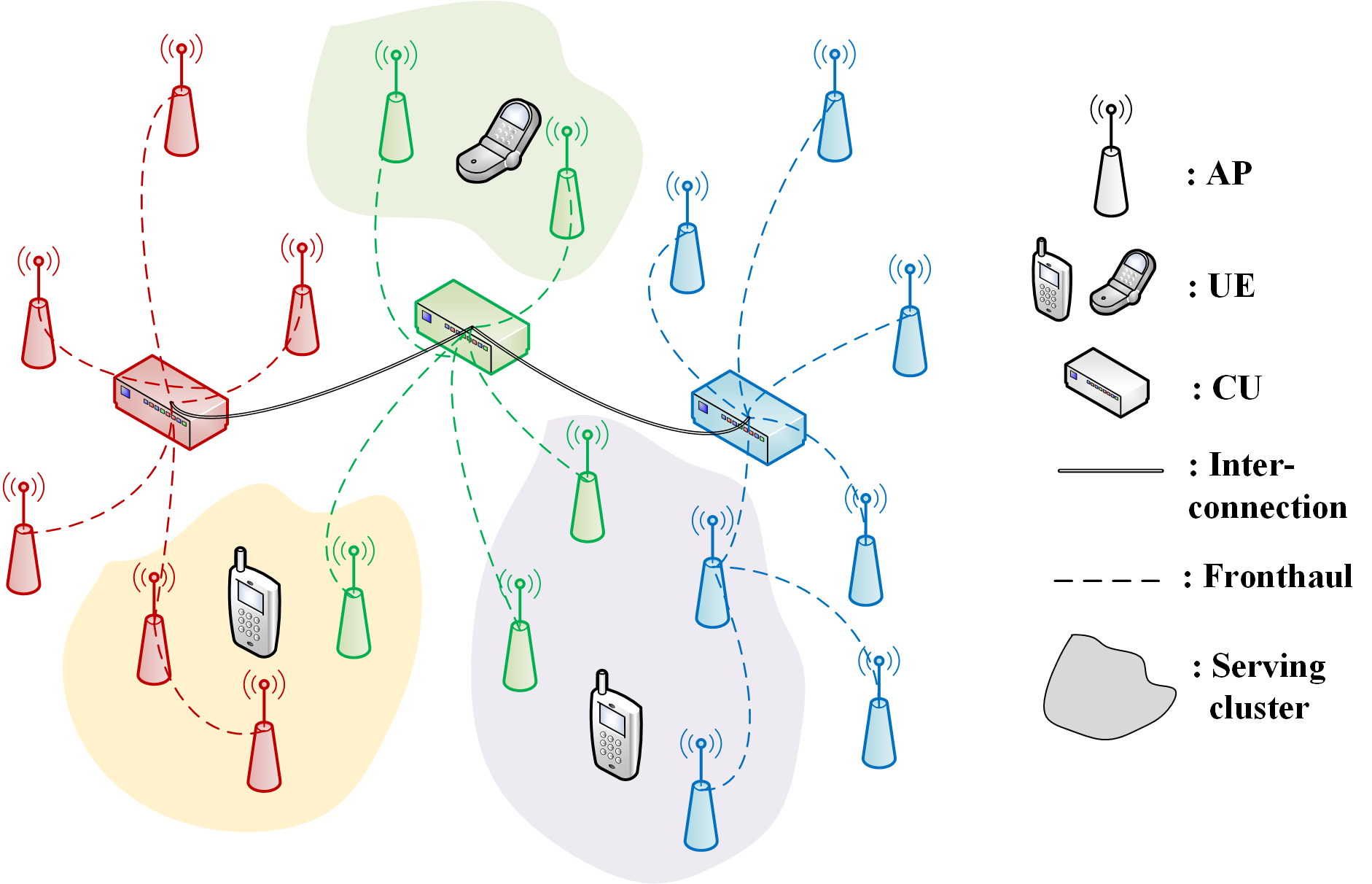}
\caption{An illustration of a user-centric cell-free network with multiple CUs.}
\label{fig_system}
\vspace{-12pt}
\end{figure}

 We consider a downlink user-centric cell-free MIMO network that comprises 
 $I$ APs and $K$ UEs, denoted by indices $\mathcal{I}=\{1, \dots, I\}$ and ${\mathcal{U}}=\{1, \dots, K\}$, respectively. 
Each AP $i$ is equipped with $M_i$ transmit antennas, serving a set of UEs ${\mathcal{U}}_i \subseteq \mathcal{U}$. 
Each UE $k$ is equipped with $N_k$ receive antennas and is jointly served by APs ${\mathcal{I}_k} \subseteq \mathcal{I}$.
As illustrated in Fig.~\ref{fig_system}, geographically distributed APs collaborate with each other either through CUs or nearby APs.

 \subsection{Non-Coherent Joint Transmission Model}
We consider that the APs cooperate in a non-coherent joint transmission mode to relieve the burden of strict synchronization among the CUs and APs.
Specifically, each UE $k$ receives different signals $\mathbf{s}_{i,k}\in \mathbb{C}^{D_{i,k}\times 1}$ from all the AP $i$ with  $D_{i,k}$ being the number of transmit streams.  
Note that only AP $i \in \mathcal{I}_k$ transmits useful signals to UE $k$.
Denote the channel matrix between AP $i$ and UE $k$ as $\mathbf{H}_{i,k} \in \mathbb{C}^{N_k \times M_i}$. 
The received signal at UE $k$ is given by
\begin{equation}\label{MU_beamforming_receive_signal}
{\by_k} = \underbrace{\sum\limits_{i \in {{\cal I}_k}} {{\bf{H}}_{i,k}{{\bf{P}}_{i,k}}{{\bf{s}}_{i,k}}}}_{\text{useful signals}}  + \underbrace{\sum\limits_{l \in {\cal U}_{-k}} {\sum\limits_{j \in {{\cal I}_l}} {{\bf{H}}_{j,k}{{\bf{P}}_{j,l}}{{\bf{s}}_{j,l}}} }}_{\text{inter-user interference}}  + {{{\bf{z}}_k}}
\end{equation}
where $\mathbf{P}_{i,k} \in \mathbb{C}^{M_i \times D_{i,k}}$ is the beamforming matrix between AP $i$ and UE $k$, $\mathbf{z}_k\sim\mathcal{CN}\left(\mathbf{0},{\sigma_k}^{2}\mathbf{I}\right)$ is the additive Gaussian noise at UE $k$.  
Note that the first term in~\eqref{MU_beamforming_receive_signal} represents the useful signals transmitted from APs $\mathcal{I}_k$ to UE $k$, while the second term is the inter-user interference.

For NCJT, successive interference cancellation (SIC) is usually adopted at the UEs to detect useful signals from their serving APs \cite{8482453,vu2020noncoherent}. 
With SIC, the achievable data rate of UE $k$ is 
\begin{equation}\label{R_k}
R_k ={\log \det \left( {{\bf{I}} + \left(\sum\limits_{i \in {{\cal I}_k}} {{\bf{H}}_{i,k}^{}{{\bf{P}}_{i,k}}{\bf{P}}_{i,k}^H{\bf{H}}_{i,k}^H}\right)  \mathbf{N}_k^{ - 1}} \right)}
\end{equation}
where  $\mathbf{N}_k \in \mathbb{C}^{N_k \times N_k}$ is the interference-plus-noise term
\begin{equation}\label{Nkk}
	{{\bf{N}}_k =}   \sum\limits_{l \in {\cal U}_{-k}} {\sum_{j \in {{\cal I}_l}} {{\bf{H}}_{j,k}{{\bf{P}}_{j,l}}} {{\bf{P}}_{j,l}^H} {\bf{H}}_{j,k}^H }+ \sigma _k^2{\bf{I}}.
\end{equation}
\vspace{-20pt}
\subsection{Problem Formulation}
We aim to jointly optimize downlink beamforming and stream allocation to maximize the WSR of the user-centric cell-free MIMO networks with NCJT,
subject to both individual power budgets at the APs and data stream limits at the UEs. 
The optimization problem is formulated as
\begin{subequations}\label{WSR_problem}
\hspace{-3mm}
    \begin{align}
    \hspace{-3mm}
\underset{\substack{\left\{ {{{\bf{P}}_{i,k}}} \right\}, \\ \{D_{i,k}\}}}{\max}
   & \sum_{k=1}^K\! \alpha_k {\log \det \! \left(\! {{\bf{I}}\! +\!\! \left(\sum\limits_{i \in {{\cal I}_k}}\!\! {{\bf{H}}_{i,k}^{}{{\bf{P}}_{i,k}}{\bf{P}}_{i,k}^H{\bf{H}}_{i,k}^H}\!\!\right)\!  \mathbf{N}_k^{ - 1}}\!\! \right)}   \\
        \operatorname{ s.t. } \  \       &  \sum\limits_{k\in \mathcal{U}_i}\operatorname{Tr}\left( \mathbf{P}_{i,k} \mathbf{P}_{i,k}^{H}\right)\leq P _{\max,i}, \ \forall i,\label{cons_p}\\
        & D^k \le N_k, \ \forall k \label{stream_a} 
    \end{align}
\end{subequations}
where $\alpha_k >0$ is the weight of the data rate on UE $k$, $P_{\max,i}$ is the power budget of AP $i$, and $D^k = \sum_{i \in \mathcal{I}_k} D_{i,k}$ denotes the total number of receive streams at UE $k$.

The joint beamforming design and stream allocation problem \eqref{WSR_problem} is non-convex and mixed-integer in nature, which is known to be NP-hard.
Furthermore, each beamforming matrix $\mathbf{P}_{i,k}$ is tightly coupled with the number of data streams $D_{i,k}$, since it determines the dimension of $\mathbf{P}_{i,k}$. 
These bring new challenges to the algorithm design to decouple the optimization variables $\{\mathbf{P}_{i,k}\}$, $\{D_{i,k}\}$, and solve \eqref{WSR_problem} with low computational complexity.

We note here that existing works \cite{8482453,vu2020noncoherent,9556147,ammar2021downlink} on WSR maximization for NCJT treat the pure beamforming optimization problem with fixed data streams $\{D_{i,k}\}$ as an NP-hard problem, 
and their beamforming algorithms for NCJT in cell-free networks are centralized approaches. 
This necessitates that all beamforming calculations be performed by the CUs, creating unavoidable computational complexity and communication overhead between APs and their connected CUs.
In contrast, we propose the RWMMSE beamforming algorithm in Section~\ref{sec_beamforming} that achieves lower computational complexity and less interaction, without any WSR compromise.

\section{Beamforming Design with Fixed Streams} \label{sec_beamforming}
It is challenging to directly solve the joint optimization problem \eqref{WSR_problem} due to the coupled decision variables $\mathbf{P}_{i,k}$ and $D_{i,k}$.
In this section, we first consider the fixed $D_{i,k}$ case to draw theoretical insights on the low-dimension beamforming structures.

With fixed $\{D_{i,k}\}$ in \eqref{WSR_problem}, the problem of beamforming design for WSR maximization subject to per-AP power constraints can be reduced to
\begin{equation}\label{WSR_max_MU}
    \begin{aligned}
        \max_{\{\mathbf{P}_{i,k}\}} \ & \sum\limits_{k = 1}^K \alpha_k {\log \det\! \left(\! {{\bf{I}} \!+ \!\!\left(\sum\limits_{i \in {{\cal I}_k}}\!\! {{\bf{H}}_{i,k}{{\bf{P}}_{i,k}}{\bf{P}}_{i,k}^H{\bf{H}}_{i,k}^H}\!\!\right)\! {\bf{N}}_k^{ - 1}} \!\!\right)}  \\
        \operatorname{ s.t. } \  \      & \eqref{cons_p}.
             \end{aligned}
\end{equation}

In the following, 
we propose three beamforming algorithms for solving problem \eqref{WSR_max_MU}: 
1) centralized WMMSE algorithm, by firstly demonstrating the applicability of the classic WMMSE approach in NCJT; 
2) distributed low-interaction RWMMSE algorithm, via studying the beamformer structure for the WSR maximization problem \eqref{WSR_max_MU}; 
\footnote{Our proposed RWMMSE algorithm is distributed, in the sense that all the beamformers at the APs are \textit{locally} obtained, instead of directly obtaining from the CUs. Furthermore, the RWMMSE algorithm does not require raw CSI and beamformer exchange between the APs and CUs.}
and 3) fully distributed Local EZF method, which can be used to initialize the RWMMSE algorithm.

\vspace{-8pt}
\subsection{ Centralized WMMSE Algorithm}\label{sec_WMMSE}
Prior works \cite{8482453} and \cite{vu2020noncoherent} assume that the efficient WMMSE approach that has been widely adopted in CJT is not applicable to NCJT.
Their proposed SCA based algorithms require solving transformed SOCP problems via the interior-point method in CVX, which is of high computational cost.
We show here that, based on our unique observations of the WSR maximization problem structure, the classic WMMSE approach is still applicable for NCJT.

In essence, the classic WMMSE algorithm is proposed to reformulate a non-convex Shannon capacity to a convex weighted mean square error (MSE) minimization problem by introducing auxiliary variables \cite{5756489}. 
Complicated WSR optimization problems can then be solved by the BCD method.
We reveal the following lemma in \cite{shi2015secure} to illustrate the equivalent transformation in the WMMSE approach.
\begin{lemma}\label{lemma1}
	(WMMSE Transformation \cite[Lemma 4.1]{shi2015secure}) 
	For any ${\bf{A}}\in {\mathbb{C}}^{n\times p}$, ${\bf{B}}\in {\mathbb{C}}^{p\times m}$ and ${\bf{N}} \succ {\bf{0}}\in {\mathbb{C}}^{n\times n}$, the following transformation holds:
\begin{equation} \label{wmmse}
    \begin{aligned}
       & \log \operatorname{det} ({\bf{I}} + {\bf{A}}{\bf{B}}{\bf{B}}^H{\bf{A}}^H{\bf{N}}^{-1}) \\
      & =    \max_{\mathbf{W} \succ \mathbf{0}, {\bf{U}}} \log \operatorname{det}(\mathbf{W})-\operatorname{Tr}({\mathbf{W}} {\bf{E}}({\bf{U}},{\bf{B}}))+m 
    \end{aligned}
\end{equation}
where ${\bf{W}} \in {\mathbb{C}}^{m\times m} $ and ${\bf{U}} \in {\mathbb{C}}^{n\times m}$ are auxiliary variables, the MSE matrix ${\bf{E}} \in {\mathbb{C}}^{m\times m}$ is defined as
\begin{equation}\label{formE}
    \mathbf{E}(\mathbf{U}, \mathbf{B}) \triangleq\left(\mathbf{I}-\mathbf{U}^{H} \mathbf{AB}\right)\left(\mathbf{I}-\mathbf{U}^{H} \mathbf{AB}\right)^{H}+\mathbf{U}^{H} \mathbf{N} \mathbf{U}.
\end{equation}
The optimal ${\bf{U}}^*$ and ${\bf{W}}^*$ for 
 \eqref{wmmse} are given by
\begin{equation} \label{formU}
\setlength{\abovedisplayskip}{2pt}
        {\mathbf{U}}^* = \left(\mathbf{N}+\mathbf{AB} \mathbf{B}^{H} \mathbf{A}^{H}\right)^{-1} \mathbf{AB},
\end{equation}
and
\begin{equation} \label{formW}
\begin{aligned}
\setlength{\abovedisplayskip}{1pt}
        {\bf{W}}^* = \left(\mathbf{I}-{\mathbf{U}}^{H} \mathbf{A} \mathbf{B} \right)^{-1}.
\end{aligned}
\end{equation}
\end{lemma}

\begin{remark}
Prior works assume that the WMMSE approach is not applicable to the WSR maximization problem in NCJT with the unique data rate structure of $
\log  \left( 1+ {\mathbf{b}^H}\mathbf{A} {\mathbf{b}}{n}^{-1} \right)$, \emph{since the rank of $\mathbf{A} = \bar{\mathbf{a}}\bar{\mathbf{a}}^H $ is generally higher than one}\cite{8482453,vu2020noncoherent}. 
However, following the fact that $1+\mathbf{v}^H\mathbf{u}=\det\left(\mathbf{I}+\mathbf{u}\mathbf{v}^H\right)$, 
the above data rate structure can be subtly rewritten as $\log \det \left( \mathbf{I}+ \bar{\mathbf{a}}^H{\mathbf{b}}{\mathbf{b}^H}\bar{\mathbf{a}}{n}^{-1} \right)$.
The WMMSE transformation in \eqref{wmmse} of \textbf{Lemma}~\ref{lemma1} can then be applied as follows.
\end{remark}

For ease of illustration, we first define the concatenation of the channel matrices and beamformers from APs $\mathcal{I}_k$ to UE $k$ as 
\begin{equation}\label{H_k}
\setlength{\abovedisplayskip}{1pt}
    {{\bf{H}}_k} \triangleq  {\left[ {{\bf{H}}_{{i_1},k}^{},{\bf{H}}_{{i_2},k}^{}, \ldots ,{\bf{H}}_{{i_{\left| {{{\cal I}_k}} \right|}},k}^{}} \right]} \in \mathbb{C}^{N_k \times M^k},
    \setlength{\belowdisplayskip}{1pt}
\end{equation}
and
\begin{equation} \label{P_k}
    {{\bf{P}}_k} \triangleq {\rm{blkdiag}}\left( {{\bf{P}}_{{i_1},k}^{},{\bf{P}}_{{i_2},k}^{}, \ldots ,{\bf{P}}_{{i_{\left| {{{\cal I}_k}} \right|}},k}^{}} \right) \in \mathbb{C}^{M^k \times D^k}
    \setlength{\belowdisplayskip}{5pt}
\end{equation}
where $M^k = \sum_{i \in \mathcal{I}_k} M_i$, and ${\left| {{{\cal I}_k}} \right|}$ is the number of serving APs $\mathcal{I}_k$ to UE $k$.
Then problem \eqref{WSR_max_MU} can be rewritten to yield the structure of the WMMSE transformation as  
\begin{equation}\label{WSR_max_MU_simplified}
    \begin{aligned}
\max_{\{\mathbf{P}_{i,k}\}} \  \  & \sum\limits_{k = 1}^K \alpha_k {\log \det \left( {\bf{I}} + {\bf{H}}_k {\bf{P}}_k {{\bf{P}}_k^H} {{\bf{H}}_k^H} {{\mathbf{N}}_k^{-1}} \right)}  \\
        \operatorname{ s.t. } \ \  \      &  \eqref{cons_p}.
    \end{aligned}
\end{equation}

Following the WMMSE transformation in \eqref{wmmse} in \textbf{Lemma}~\ref{lemma1}, problem \eqref{WSR_max_MU_simplified} can be equivalently transformed to
\begin{equation}\label{WSR_max_MU_WMMSE_trans}
    \begin{aligned}
  \underset{\substack{\left\{ {{{\bf{P}}_{i,k}}} \right\},{\{\bf{U}}_k\} \\ \{{\bf{W}}_k \succ {\bf{0}}\}}}{\max}
 \ & \sum\limits_{k = 1}^K\! \alpha_k \! \left({\log \det \! \left( {{{\bf{W}}_k}} \right)\!\! -\! \!{\rm{Tr}}\left( {{{\bf{W}}_k}{{\bf{E}}_k} \!\left( {\bf{U}}_k, \!{\bf{P}} \right)} \right)} \right) \\
    \operatorname{ s.t. } ~ \ \ \   \     & \eqref{cons_p}
    \end{aligned}
    \setlength{\belowdisplayskip}{1pt}
\end{equation}
where ${\bf{P}}=\left\{ {{{\bf{P}}_{i,k}}} \right\}$, ${\bf{U}}_k \in \mathbb{C}^{N_k \times D^k}$ and ${\bf{W}}_k \succ \mathbf{0} \in \mathbb{C}^{D^k \times D^k}$ are auxiliary variables, the MSE matrix ${{\bf{E}}_k} \in \mathbb{C}^{D^k \times D^k}$ is defined as
\begin{equation} \label{def_E_P}
\begin{aligned}
\setlength{\abovedisplayskip}{2pt}
	    {{\bf{E}}_k}\left( {\bf{U}}_k,{\bf{P}} \right) \triangleq &\left( {{\bf{I}} - {\bf{U}}_k^H{{\bf{H}}_k}{{\bf{P}}_k}} \right){\left( {{\bf{I}} - {\bf{U}}_k^H{{\bf{H}}_k}{{\bf{P}}_k}} \right)^H}\\
	    & + {\bf{U}}_k^H{{\bf{N}}_k}{\bf{U}}_k.
\end{aligned}
\end{equation}

Note that problem \eqref{WSR_max_MU_WMMSE_trans} is convex with respect to (w.r.t.) $\{{\bf{U}}_k\}$, $\{{\bf{W}}_k\}$ and $
\{\mathbf{P}_{i,k}\}$, respectively.
From \eqref{formU} and \eqref{formW} in \textbf{Lemma}~\ref{lemma1}, the optimal $\{{\bf{U}}_k^*\}$ and $\{{\bf{W}}_k^*\}$ with fixed $\{\mathbf{P}_{i,k}\}$ can be updated by
\begin{equation}\label{MU_U_MMSE}
\setlength{\abovedisplayskip}{2pt}
    {\bf{U}}_k^* = {\left( {{{\bf{N}}_k} + {{\bf{H}}_k}{{\bf{P}}_k}{\bf{P}}_k^H{\bf{H}}_k^H} \right)^{ - 1}}{{\bf{H}}_k}{{\bf{P}}_k}, \ \forall k ,
\end{equation}
and
\begin{equation}\label{MU_W}
\setlength{\abovedisplayskip}{1pt}
{\bf{W}}_k^* =  {\left( {{\bf{I}} - {{\left( {{\bf{U}}_k^*} \right)}^{H}}{{\bf{H}}_k}{{\bf{P}}_k}} \right)^{ - 1}} ,\ \forall  k .
\end{equation}
Fixing $\{{\bf{U}}_k^*\}$ and $\{{{\bf{W}}}_k^*\}$, the WMMSE update of $\{{{\bf{P}}}_{i,k}^*\}$ is given by 
\begin{equation}\label{WMMSE_P}
\begin{aligned}
\setlength{\abovedisplayskip}{1pt}
{\bf{P}}_{i,k}^{*} \! = \! {\left( {\sum_{l \in \mathcal{U}}\! { \alpha_l{{\bf{H}}_{i,l}^H{{\bf{A}}_l}{{\bf{H}}_{i,l}}} \! + \! {\mu _i}{\bf{I}}} }\!\! \right)^{ - 1}} \!\!\!\!\alpha_k{\bf{H}}_{i,k}^H\!{\bf{U}}_k^*{\bf{W}}_k^*{\bf{\Xi }}_{i,k}^H
\setlength{\belowdisplayskip}{2pt}
\end{aligned}
\end{equation}
where $\mu_i \geq 0$ is a Lagrangian constant for the maximum transmit power constraint \eqref{cons_p}. 
In \textbf{Algorithm}~\ref{al:WMMSE_MU}, we summarize our proposed WMMSE algorithm, which requires centralized computation of $\{{{\bf{P}}}_{i,k}^*\}$ at the CUs.
\begin{remark} \label{remark1}
For WSR maximization in NCJT, the BRnB method in \cite{vu2020noncoherent} has an $\mathcal{O}\left(\left(\sum_{i \in \mathcal{I}}{M_i^2}\right)^3\right)$ computational complexity. 
The SCA based algorithm in \cite{8482453} achieves a lower $\mathcal{O}\left( \left(\sum_{i \in \mathcal{I}}{M_{i} \sum_{k \in \mathcal{U}_i}D_{i,k}}\right)^3 \right)$ computational complexity.
The InAP and ADMM combined method in \cite{vu2020noncoherent} has a even lower $\mathcal{O}\left(\left(\sum_{i \in \mathcal{I}}{M_i}\right)^3\right)$ complexity. 
The FP based algorithm in~\cite{ammar2021downlink} achieves the current lowest $\mathcal{O}(\sum_{i \in \mathcal{I}}{M_i^3})$ computational complexity for the single UE receive antenna case. 
Our proposed WMMSE algorithm achieves the same $\mathcal{O}(\sum_{i \in \mathcal{I}}{M_i^3})$ computational complexity, for the more general case with multiple UE receive antennas.
\end{remark}

\setlength{\textfloatsep}{8pt}
\begin{algorithm}[!t]
    \renewcommand{\algorithmicrequire}{\textbf{Interact:}}
    \renewcommand{\algorithmicensure}{\textbf{Output:}}
    \caption{Centralized WMMSE Algorithm}  \label{al:WMMSE_MU}
    \begin{algorithmic}[1]
    \REQUIRE Each AP $i$ sends its $\{{{\bf{H}}}_{i,k}\}$ to the connected CUs;\\
    \STATE Initialize $\{{\bf{P}}_{i,k}\}$ in CUs;\\
    \STATE \textbf{Repeat}~$\leftarrow$[CUs]\\
    \STATE  \quad Update $\{{\bf{U}}_k^*\}$ via \eqref{MU_U_MMSE};
    \STATE  \quad Update $\{{\bf{W}}_k^*\}$ via \eqref{MU_W};\\
    \STATE  \quad Update $\{{\bf{P}}_{i,k}^*\}$ via \eqref{WMMSE_P};\\
    \STATE  \textbf{Until}~Converge\\
    \REQUIRE  CUs transmit $\{{\bf{P}}_{i,k}^*, k \in \mathcal{U}_i \}$ to each AP $i$.
    \end{algorithmic}
    \setlength{\belowdisplayskip}{5pt}
\end{algorithm}

\vspace{-15pt}
\subsection{Distributed Low-Interaction RWMMSE Algorithm}

The current lowest $\mathcal{O}(\sum_{i \in \mathcal{I}}{M_i^3})$ computational complexity of our proposed WMMSE algorithm can still be high, especially for MIMO systems with massive transmit antennas.
Moreover, the centralized WMMSE algorithm requires each AP $i$ to communicate the channel matrix $\{{{\bf{H}}}_{i,k}, k \in \mathcal{U}_i\}$ to the CUs, and the CUs to communicate the beamforming matrix $\{{\bf{P}}_{i,k}^*, k \in \mathcal{U}_i \}$ back to each AP $i$, causing high interaction between the APs and the CUs.
These motivate us to explore low-complexity and low-interaction beamforming algorithms. 

In the following, we draw some theoretical insights on the structures of the BCD solutions to problem \eqref{WSR_max_MU}, which will be leveraged later to further reduce the algorithm complexity and interaction.

\subsubsection{Properties of Beamforming for WSR Maximizatoin}
We now look through the inherent properties of the WSR maximization problem \eqref{WSR_max_MU} to study the structure of ${\bf{P}}_{i,k}^{*}$. 
For ease of expression, we concat the channel matrices between AP $i$ and all the UEs as follows
\begin{equation}
	\bar{\bf{H}}_i \triangleq \left[ {\bf{H}}_{i,1}^H,\dots,{\bf{H}}_{i,K}^H\right]^H\in \mathbb{C}^{\sum_{k \in \mathcal{U}}N_k \times M_i}.
\end{equation}

The following \textbf{Theorem}~\ref{theorem_sum} shows if $\bar{\bf{H}}_i$ is of full column rank, the beamformer ${\bf{P}}_{i,k}^{*}$ for WSR maximization consumes full transmit power.
\begin{theorem} \label{theorem_sum}(Full Power Property):
For any AP $i$, if $M_i \geq \sum_{k \in \mathcal{U}}N_k$ and $\bar{\bf{H}}_i^H$ consists of $\sum_{k \in \mathcal{U}}N_k$ linearly independent vectors of size $M_i \times 1$,
 the local optimal beamformer ${\mathbf{P}}_{i,k}^*$ always satisfies the full transmit power constraint, i.e. $\sum_{k\in \mathcal{U}_i}
\left\| {\mathbf{P}}_{i,k}^* \right\|_{\text{F}}^2 =  P_{\max,i }$.
\end{theorem}
\begin{IEEEproof}
See Appendix \ref{Appendix0}.
\end{IEEEproof}
The following proposition shows another property on the power consumption of the beamformer ${\mathbf{P}}_{i,k}^*$, i.e., the column vectors of ${\mathbf{P}}_{i,k}^*$ that are orthogonal to the channel matrix ${\bar{\mathbf{H}}_i}$ may consume power but do not contribute to the WSR.

\begin{Proposition}\label{Pro_1}(Null Space Property):
    The part of the beamformer ${{\bf{P}}_{i,k}^*}$ in the null space of ${\bar{\mathbf{H}}_i}^{H}$ consumes power but does not contributes to the WSR.
\end{Proposition}
\begin{IEEEproof}
See Appendix \ref{Appendix_proof_proposition}.
\end{IEEEproof}

The following theorem shows that each beamformer ${{\bf{P}}_{i,k}^*}$ for WSR maximization has a low-dimensional substitution.
\begin{theorem} \label{theom2} (Low-Dimension Substitution):
	Any local optimal solution $\{{\bf{P}}_{i,k}^{*} \in \mathbb{C}^{M_i \times D_{i,k}} , k \in \mathcal{U}_i\}$ to problem \eqref{WSR_max_MU_simplified} must lie in the column space of $\bar{{\bf{H}}}_i^H$, i.e.,
	\begin{equation}\label{MU_precoder_structure}
   \mathbf{{P}}_{i,k}^{*}={\bar{{\bf{H}}}_i}^H\mathbf{X}_{i,k}^*,\forall k 
    \end{equation}
    where $\mathbf{X}_{i,k}^* \in \mathbb{C}^{\sum_{k \in \mathcal{U}} N_k \times D_{i,k}}$ is a low-dimensional substitution of $\mathbf{P}_{i,k}^*$. 
\end{theorem}
\begin{IEEEproof}
See Appendix~\ref{Appendix_proof_X}.
\end{IEEEproof}
\begin{remark} (Low-Complexity and Low-Interaction Properties):
By investigating the structure of the beamformer $\mathbf{{P}}_{i,k}^{*}$ in \eqref{MU_precoder_structure} of \textbf{Theorem}~\ref{theom2}, solution to ${\mathbf{P}}_{i,k}^{\text{*}}$ in \eqref{WMMSE_P} with the size of ${M_i \times D_{i,k}}$ can be equivalently reduced to $\mathbf{X}_{i,k}^*$ with the size of $ {\sum_{k \in \mathcal{U}} N_k \times D_{i,k}}$. 
In addition, the interaction between the APs and their connected CUs is changed from $\mathbf{{P}}_{i,k}^{*}$ to $\mathbf{{X}}_{i,k}^{*}$.
By solving for $\mathbf{{X}}_{i,k}^{*}$ instead of $\mathbf{{P}}_{i,k}^{*}$, both the computation cost and the communication overhead can be substantially reduced,
especially for the networks with a large number of transmit antennas $M_i \gg \sum_{k \in \mathcal{U}} N_k$.
\end{remark}
\begin{remark}
The full power and the low-dimension subspace properties presented in \textbf{Theorem}~\ref{theorem_sum} and  \textbf{Theorem}~\ref{theom2} are substantially different from the results in \cite{zhao2023rethinking} in the following aspects. 
First, the findings in \cite{zhao2023rethinking} are limited to CJT, which cannot be directly applied to NCJT in this work.
Second, the results in \cite{zhao2023rethinking} are limited to the \emph{single}-AP case, while we study the more general cell-free networks with \emph{multiple} APs.
Note that the proof of the full power property in \cite{zhao2023rethinking} relies on the contradiction of one single Lagrangian multiplier in the Karush-Kuhn-Tucker (KKT) conditions, which cannot be directly applied to the case of multiple APs.
Finally, our proofs based on the linear independence of columns in  $\bar{\bf{H}}_i^H$ avoids the tedious discussion on the KKT conditions in \cite{zhao2023rethinking}.
\end{remark}

\subsubsection{Problem Reformulation}
 We now utilize the low-dimensional substitution property in \textbf{Theorem}~\ref{theom2} to reformulate problem \eqref{WSR_max_MU_simplified}. 
 From the low-dimension substitution \eqref{MU_precoder_structure} and the definition of ${\mathbf{P}}_k$ in \eqref{P_k}, we have
\begin{equation}\label{R_WMMSE_P_k}
    {\mathbf{P}}_k = {{\tilde{{\bf{H}}}_{{\mathcal{I}}_k}}^H{\tilde{{\bf{X}}}_{{\mathcal{I}}_k}}}
\end{equation}
where ${\tilde{{\bf{H}}}_{{\mathcal{I}}_k}} \triangleq \textrm{blkdiag}\left(\! {\bar{{\bf{H}}}}_{i_1},\dots,{\bar{{\bf{H}}}}_{i_{\left| {{\mathcal{I}}_k} \right|}}\right)\!\in\! \mathbb{C}^{\left| {\mathcal{I}_k} \right | \sum_{k\in \mathcal{U}}  N_k \times M^k}$ 
and ${\tilde{{\bf{X}}}_{{\mathcal{I}}_k}}\triangleq \textrm{blkdiag}\left(\! {{{\bf{X}}}}_{i_1,k},\dots,{{{\bf{X}}}}_{i_{\left| {{\mathcal{I}}_k} \right|},k}\right)\! \in \! \mathbb{C}^{\left| {\mathcal{I}_k} \right | \sum_{k\in \mathcal{U}}  N_k \times D^k}$ 
are the concatenations of the channel matrices $\{\bar{\mathbf{H}}_i , i\in\mathcal{I}_k \}$  and the low-dimension substitutions $\{\mathbf{X}_{i,k}, i\in\mathcal{I}_k\}$.

%
%
Substituting \eqref{R_WMMSE_P_k} into the WSR maximization problem \eqref{WSR_max_MU_simplified}, we have a low-dimension substitution problem given by
\begin{subequations}\label{WSR_max_MU_simplified_X}
    \begin{align}
   \max_{\{\!\mathbf{X}_{i,k}\!\}} \  & \sum\limits_{k\in\mathcal{U}} \!\alpha_k {\log \det\! \left(\! {\bf{I}}\!+ \!{\bf{H}}_k {{\tilde{{\bf{H}}}_{{\mathcal{I}}_k}}^H{\tilde{{\bf{X}}}_{{\mathcal{I}}_k}}}\!{\tilde{{\bf{X}}}_{{\mathcal{I}}_k}^H}\!{\tilde{{\bf{H}}}_{{\mathcal{I}}_k}}\!{{\bf{H}}_k^H} {{\mathbf{N}}_k^{ \!-\!1}}\! \right)}  \\
        \operatorname{ s.t. } \   \      &  \sum\limits_{k\in \mathcal{U}_i}\!\!\operatorname{Tr}\left( {\bar{{\bf{H}}}_i}^H\mathbf{X}_{i,k} {\mathbf{X}_{i,k}^H{\bar{{\bf{H}}}_i}}\right)\leq P_{\max,i}, \ \forall i. \label{cons_x}
    \end{align}
\end{subequations}
Following the WMMSE transformation in \textbf{Lemma}~\ref{lemma1}, problem~\eqref{WSR_max_MU_simplified_X} can be equivalently transformed to
\begin{equation}\label{WSR_max_MU_RWMMSE_trans}
    \begin{aligned}
          \underset{\substack{\left\{ {{{\bf{X}}_{i,k}}} \right\},{\{\!\bf{U}}_k\!\} \\ \{{\bf{W}}_k \succ {\bf{0}}\}}}{\max} \
    & \sum\limits_{k\in\mathcal{U}}\! \alpha_k \! \left({\log \det \! \left( {{{\bf{W}}_k}} \right)\!\! -\! \!{\rm{Tr}}\left( {{{\bf{W}}_k}{{\bf{E}}_k} \!\left( {\bf{U}}_k, \!{\bf{X}} \right)} \right)} \right) \\
         \operatorname{ s.t. } ~ \ \ \         & \eqref{cons_x}    \end{aligned}
\end{equation}
where ${\bf{X}}\triangleq\{{\bf{X}}_{i,k}\}$ is the set of low-dimension beamformer substitution, and the MSE matrix is defined as
\begin{equation} \label{def_E_X}
\begin{aligned}
	    {{\bf{E}}_k}\!\left(\! {\bf{U}}_k,\!{\bf{X}} \right) 
	  \! \triangleq \!
	   & \left( {{\bf{I}}\!\!- \!\!{\bf{U}}_k^H\!{{\bf{H}}_k}{{\tilde{{\bf{H}}}_{{\mathcal{I}}_k}}^H\!{\tilde{{\bf{X}}}_{{\mathcal{I}}_k}}}} \!\right)\!\!
	   {\left( {{\bf{I}}\!\! -\!\! {\bf{U}}_k^H\!{{\bf{H}}_k}{{\tilde{{\bf{H}}}_{{\mathcal{I}}_k}}^H\!{\tilde{{\bf{X}}}_{{\mathcal{I}}_k}}}}
	   \!\! \right)\!^H}\\
	    & + {\bf{U}}_k^H{{\bf{N}}_k}{\bf{U}}_k.
\end{aligned}
\end{equation}

Similar to the BCD updates in \eqref{MU_U_MMSE} and \eqref{MU_W}, with fixed $\mathbf{X}$ we update $\{{\mathbf{U}}_k^*\}$  and $\{{\mathbf{W}}_k^*\}$ by replacing ${{\bf{P}}_k}$ with its low-dimension substitution in \eqref{R_WMMSE_P_k} as
\begin{equation}\label{MU_U_MMSE_rwmmse}
\hspace{-1mm}
\begin{aligned}
	    {\bf{U}}_k^*\!\! = \!& {\left( {{{\bf{N}}_k}\!\! +\!\! {\bf{H}}_k {{\tilde{{\bf{H}}}_{{\mathcal{I}}_k}}^H\!{\tilde{{\bf{X}}}_{{\mathcal{I}}_k}}}\!{\tilde{{\bf{X}}}_{{\mathcal{I}}_k}^H}\!{\tilde{{\bf{H}}}_{{\mathcal{I}}_k}}\!{{\bf{H}}_k^H}}\! \right)\!\!^{\!-\! 1}}\! {{\bf{H}}_k}\!{{\tilde{{\bf{H}}}_{{\mathcal{I}}_k}}^H\!{\tilde{{\bf{X}}}_{{\mathcal{I}}_k}}},\!\!
	     \  \forall k,
\end{aligned}
\end{equation}
and
\begin{equation}\label{MU_W_rwmmse}
{\bf{W}}_k^* =  {\left( {{\bf{I}} - {{\left( {{\bf{U}}_k^*} \right)}^{H}}{{\bf{H}}_k}{{\tilde{{\bf{H}}}_{{\mathcal{I}}_k}}^H{\tilde{{\bf{X}}}_{{\mathcal{I}}_k}}}} \right)^{ - 1}} ,\ \forall  k.
\end{equation}
Fixing $\{{\mathbf{U}}_k^*\}$ and $\{{\mathbf{W}}_k^*\}$, optimization problem \eqref{WSR_max_MU_RWMMSE_trans} can be reduced to
\begin{equation}\label{RWMMSE_Probelm_X}
    \begin{aligned}
        \mathop {\min }\limits_{\left\{ {{{\bf{X}}_{i,k}}} \right\}}   \ &  \sum\limits_{k\in\mathcal{U}} 
    \alpha_k {{\rm{Tr}}\left( {{{\bf{W}}_k}{{\bf{E}}_k}
    \left( {\bf{U}}_k^*,{\bf{X}} \right)} \right)}   \\
        \operatorname{ s.t. } \  \      &  \ \eqref{cons_x}.
    \end{aligned}
\end{equation}
Dropping the content terms in the objective of problem~\eqref{RWMMSE_Probelm_X}, we have \begin{equation}
\hspace{-0.8mm}
\label{RWMMSE_Probelm_X1}
\begin{aligned}
\min_{\{\!{\bf{X}}_{i,k}\!\}}
&\sum\limits_{k\in\mathcal{U}} \alpha_k\sum\limits_{i \in \mathcal{I}_k} {\rm{Tr}}\left( {\mathbf{X}_{i,k}^H{\bar{{\bf{H}}}_i}{\bf{H}}_{i,k}^H{{\bf{A}}_k}{{\bf{H}}_{i,k}}{\bar{{\bf{H}}}_i}^H\mathbf{X}_{i,k}} \right) \\
& -\!\! 2\!\sum\limits_{k\in\mathcal{U}}\!\! \alpha_k\!\! \sum\limits_{i \in \mathcal{I}_k}\!\!{\mathop{\rm Re}\nolimits}\! \left\{ {{\rm{Tr}}\!\left( {{{\bf{\Xi }}_{i,k}}\!{{\bf{W}}_k^*}(\!{\bf{U}}_k^*)\!^H\!{{\bf{H}}_{i,k}}{\bar{{\bf{H}}}_i}^H\!\mathbf{X}_{i,k}}\! \right)}\! \right\} \!\!\!\!\! \\
&+\!\! \sum\limits_{k\in\mathcal{U}}\!\! \alpha_k\!{{\rm{Tr}}\!\!\left(\!\! {{{\bf{A}}_k}\!\!\!\sum\limits_{j \in {\mathcal{I}_l}} {\sum\limits_{l \in \mathcal{U}_{-k}}\!\!\! {{{\bf{H}}_{j,k}}\!{{\bar{{\bf{H}}}_j}^H\!\mathbf{X}_{j,l}}{{\bar{{\bf{H}}}_j}^H\mathbf{X}_{j,l}^H}{\bf{H}}_{j,k}^H} } } \!\!\right)}\\
\operatorname{ s.t. }   \ & \ \eqref{cons_x} 
\end{aligned}
\end{equation}
where ${\bf{A}}_k\triangleq {\bf{U}}_k^{}{{\bf{W}}_k}{\bf{U}}_k^H \in \mathbb{C}^{N_k \times N_k}$ and 
${\bf{\Xi}}_{i,k} \in {\mathbb{B}}^{D_{i,k} \times D^k}$ is a binary matrix with the $\left({i - 1} \right)\times {D_{i,k}} + 1$ to $i \times D_{i,k}$ columns of ${\bf{\Xi}}_{i,k}$ being $\mathbf{I}_{D_{i,k}}$ and the rest of elements being zero when $i \in \mathcal{I}_k$, otherwise ${\bf{\Xi}}_{i,k} = \bf{0}$ for $i \notin \mathcal{I}_k$.

We note here that the updates of ${\{{\bf{X}}_{i,k}\}}$ in \eqref{RWMMSE_Probelm_X1} can be decoupled across APs, each AP $i$ solves its own optimization problem
\begin{equation}\label{RWMMSE_Probelm_X_i_k}
    \begin{aligned}
    \hspace{-2mm}
\min_{\{{\bX_{i,k}}, k \in \mathcal{U}_i\}} &\! \sum\limits_{l \in \mathcal{U}} \sum\limits_{k \in \mathcal{U}_i} \alpha_l {{\rm{Tr}}\left( { { {{{\bf{A}}_l}{{\bf{H}}_{i,l}}{{\bar{{\bf{H}}}_i}^H\mathbf{X}_{i,k}}{\mathbf{X}_{i,k}^H{\bar{{\bf{H}}}_i}}{\bf{H}}_{i,l}^H} } } \right)}  \\
& \!-\!\! 2\!\!\sum\limits_{k \in \mathcal{U}_i}\!\!\alpha_k{\mathop{\rm Re}\nolimits}\! \left\{\! {{\rm{Tr}}\!\left( {{{\bf{\Xi }}_i}\!{{\bf{W}}_k^*}\!({\bf{U}}_k^*)\!^H\!{{\bf{H}}_{i,k}}{\bar{{\bf{H}}}_i}^H\!\mathbf{X}_{i,k}} \right)}\! \right\} \\
        \operatorname{ s.t. } \ \ \ \      &  \eqref{cons_x}.
    \end{aligned}
\end{equation}
Different from the sequential beamforming updates in \cite{8482453} and \cite{9556147}, the APs can update their own beamformer ${\bf{P}}_{i,k}$ (WMMSE) or ${\bf{X}}_{i,k}$ (RWMMSE) by solving problem \eqref{RWMMSE_Probelm_X_i_k} in parallel.

Problem \eqref{RWMMSE_Probelm_X_i_k} is convex quadratic w.r.t. ${\bf{X}}_{i,k}$.
Using the Lagrangian multiplier method, the optimal solution $\{{\bf{X}}_{i,k}^*\}$ to problem \eqref{RWMMSE_Probelm_X_i_k} is given by
\begin{equation}\label{RWMMSE_X}
\setlength{\abovedisplayskip}{1pt}
\setlength{\belowdisplayskip}{5pt}
\begin{aligned}
        {\bf{X}}_{i,k}^*  = & {\left( {\sum\limits_{l \in \mathcal{U}} \alpha_l{ {\bar{{\bf{H}}}_i{\bf{H}}_{i,l}^H{{\bf{A}}_l}{{\bf{H}}_{i,l}}\bar{{\bf{H}}}_i^H}  + {\lambda _i}{\bar{{\bf{H}}}_i\bar{{\bf{H}}}_i^H}} } \right)^{ - 1}}\\
        &\times \alpha_k \bar{{\bf{H}}}_i{\bf{H}}_{i,k}^H{\bf{U}}_k^{*}{\bf{W}}_k^*{\bf{\Xi }}_{i,k}^H
\end{aligned}
\end{equation}
where $\lambda_i \geq 0$ is the Lagrangian multiplier that can also be obtained by the bisection method. 
The detailed RWMMSE process is described in \textbf{Algorithm}~\ref{al:RWMMSE_MU}.
\begin{remark}(Low-Interaction Execution of the RWMMSE Algorithm):
When executing the RWMMSE algorithm, each AP $i$ first transmits $\{\bar{{\bf{H}}}_i\bar{{\bf{H}}}_i^H \! \!\in \!\mathbb{C}^{ \sum_{k \in \mathcal{U}}\!N_k \times \sum_{k \in \mathcal{U}}\!N_k }\}$ (with lower dimension than the interaction $\{\bar{{\bf{H}}}_i\}$ in the WMMSE algorithm) to the connected CUs. 
The CUs then transmit $\{{\bf{X}}_{i,k}^*\!\!\in\!\! \mathbb{C}^{ M_i  \times D_{i,k} }\!, k \! \in \! \! \mathcal{U}_i \}$ back to their connected APs, while the WMMSE approach need to communicate complete beamformer $\{{\bf{P}}_{i,k}^*,k \! \in \! \mathcal{U}_i\}$ with higher dimension. 
Finally, each AP~$i$ calculates their beamforming matrices via~\eqref{MU_precoder_structure}~locally. 
\end{remark}
\begin{remark}\label{WMMSE_to_RWMMSE}
The RWMMSE algorithm is equivalent to the centralized WMMSE algorithm in Section~\ref{sec_WMMSE}. 
Due to the fact that $\left({\bf{I}} + {\bf{A}}{\bf{B}} \right)^{-1}{\bf{A}} = {\bf{A}}\left({\bf{I}} + {\bf{B}}{\bf{A}} \right)^{-1}$, the WMMSE beamformer update ${\bf{P}}_{i,k}^{\text{*}}$ in \eqref{WMMSE_P} can be rewritten  to follow the low-dimension substitute property in \textbf{Theorem}~\ref{theom2}.
Specifically, the concatenation of $\{{\bf{P}}_{i,k}^{{*}},k\in\mathcal{U}\}$ can be expressed as
\begin{equation}
\begin{aligned}
	    {\bf{P}}_{i}^{\text{*}}
	    & = {\bar{\bf{H}}}_i^H \underbrace{{\bf{U}}
    \left( {\bf{\Omega }} {\bf{U}}^H {\bar{\bf{H}}}_i{\bar{\bf{H}}}_i^H{\bf{U}} + \mu_i {\bf{W}}^{-1}  \right)^{-1} {\bf{\Omega }}{\bar{\bf{\Xi }}}_i^H}_{{\bf{X}}_i} \\
    & = {\left[ { {\bf{P}}_{{i},1}^*, {\bf{P}}_{{i},2}^*, \ldots ,{\bf{P}}_{i,K}^* } \right]} \in \mathbb{C}^{M_i \times \sum_{k \in \mathcal{U}}D_{i,k}}
\end{aligned}
    \setlength{\belowdisplayskip}{5pt}
\end{equation}
where we define
${\bf{U}}\! \triangleq \!\operatorname{blkdiag} \left( {\bf{U}}_1^*,\dots ,\! {\bf{U}}_K^* \right)  \!\in\! \mathbb{C}^{\sum\limits_{k \in \mathcal{U}}\!\! N_k \times\! \sum\limits_{k \in \mathcal{U}}\!\! D^k}$,
${\bf{W}} \triangleq \operatorname{blkdiag} \left( {\bf{W}}_1^*,\dots, {\bf{W}}_K^* \right)  \in \mathbb{C}^{\sum\limits_{k \in \mathcal{U}} D^k \times \sum\limits_{k \in \mathcal{U}} D^k}$,
${\bf{\Omega} } \triangleq \operatorname{blkdiag} \left( {\bf{\alpha }}_{1}{\bf{I}
}_{D^{1}},\dots, {\bf{\alpha }}_{K}{\bf{I}
}_{D^{K}} \right) \in \mathbb{R}^{\sum\limits_{k \in \mathcal{U}} D^k \times \sum\limits_{k \in \mathcal{U}} D^k}$,
and ${\bar{\bf{\Xi }}}_i \triangleq \operatorname{blkdiag} \left( {\bf{\Xi }}_{i,1},\dots ,{\bf{\Xi }}_{i,K} \right) \in \mathbb{R}^{\sum\limits_{k \in \mathcal{U}} D_{i,k} \times \sum\limits_{k \in \mathcal{U}} D^k}$.
\end{remark}
%
%
\begin{algorithm}[t]
    \renewcommand{\algorithmicrequire}{\textbf{Interact:}}
    \renewcommand{\algorithmicensure}{\textbf{Output:}}
    \caption{Distributed RWMMSE Algorithm}  \label{al:RWMMSE_MU}
    \begin{algorithmic}[1]
    \STATE Each AP $i$ calculates $\{\bar{{\bf{H}}}_i\bar{{\bf{H}}}_i^H\}$;\\
    \REQUIRE Each AP $i$ sends $\{\bar{{\bf{H}}}_i\bar{{\bf{H}}}_i^H\}$ to the connected CUs;\\
    \STATE Initialize $\{{\bf{X}}_{i,k}\}$ in CUs;\\
    \STATE \textbf{Repeat}~$\leftarrow$[CUs]\\
    \STATE  \quad Update $\{{\bf{U}}_k^*\}$ via \eqref{MU_U_MMSE};
    \STATE  \quad Update $\{{\bf{W}}_k^*\}$ via \eqref{MU_W};\\
    \STATE  \quad Update $\{{\bf{X}}_{i,k}^*\}$ via \eqref{RWMMSE_X};\\
    \STATE  \textbf{Until}~Converge\\
    \REQUIRE  CUs transmit $\{{\bf{X}}_{i,k}^*, k\! \in \! \mathcal{U}_i \}$ to the connected AP $i$;\\
    \ENSURE Each AP $i$ calculates $\{\mathbf{P}_{i,k}^*=\bar{\mathbf{H}}_{i}^H\mathbf{X}_{i,k}^*, k \in \mathcal{U}_i \}$.
    \end{algorithmic}
\end{algorithm}
%
%
\subsection{Fully Distributed Local EZF Algorithm }
Motivated by the unique property of NCJT that signals transmitted from the serving APs of a UE are independent, we extend the EZF method for CJT that requires CSI interaction across the network\cite{zhao2023rethinking}, to a fully distributed EZF method for NCJT without any CSI interaction.
Our proposed fully distributed EZF algorithm can be used to initialize the iteration of the RWMMSE algorithm for better system performance.
\subsubsection{Local EZF Method} 
The idea of the EZF method is to perform singular value decomposition (SVD) on the channel matrix first, and then does zero-forcing (ZF) on the effective channel matrix composed of only the right singular vectors. 

Specifically, each AP $i$ performs SVD on $\{{\bf{H}}_{i,k}, k \in \mathcal{U}_i\}$ to get its right singular vectors
\begin{equation}
    {\bf{H}}_{i,k} = \bar{{\bf{U}}}_{i,k} \bar{\boldsymbol{\Sigma}}_{i,k}\bar{\bV}_{i,k}^{H}
\end{equation}
where $\bar{{\bf{U}}}_{i,k} \in \mathbb{C}^{N_k \times N_k}$ and $\bar{\bV}_{i,k} \in \mathbb{C}^{M_i \times N_k}$
are unitary matrices of the left and right singular vectors, and $\bar{\boldsymbol{\Sigma}}_{i,k} \in \mathbb{C}^{N_k \times N_k}$ is a diagonal matrix with descending singular values. 
The effective channel of AP~$i$ ${\bar{\mathbf{V}}}_{i} \in \mathbb{C}^{M_i \times \sum_{k \in \mathcal{U}_i}D_{i,k}}$ can be formed by concating the first $D_{i,k}$ columns of $\{\bar{\bV}_{i,k} , k \in \mathcal{U}_i\}$ as 
\begin{equation}\label{V_i}
 {\bar{\mathbf{V}}}_{i} \!=\!\left[ \bar{\bV}_{i,k_1}\!\left( :, 1\!\!:\!D_{i,k_1} \right), \dots,\bar{\bV}_{i,k_{\left| {{{\cal U}_i}} \right|}} \left( :, 1:D_{i,k_{\left| {{{\cal U}_i}} \right|}} \right) \right]   .
\end{equation}
Finally, each AP $i$ performs ZF and power scaling on $\bar{\bV}_i$ in parallel. The Local EZF beamformer of AP $i$ is given by
\begin{equation}\label{P_Local_EZF}
\begin{aligned}
  {\mathbf{P}}_i^{\text{Local EZF}} & =\frac{{\left({{\bar{\bf{V}}}^H_i}\right)^\dag }}{{{{\left\| {\left({{\bar {\bf{V}}}_i^H}\right)^\dag } \right\|}_\text{F}}}}\sqrt {{P_{\max ,i}}}  \\
  & = {\left[ { {\bf{P}}_{{i},k_1}, {\bf{P}}_{{i},k_2}, \dots ,{\bf{P}}_{i,k_{\left| {{{\cal U}_i}} \right|}} } \right]}, \ \forall i.
\end{aligned}
\end{equation}
We present our Local EZF method for NCJT in \textbf{Algorithm}~\ref{al:local_EZF}. 
\subsubsection{Initialization of the RWMMSE Algorithm}
We can show that the Local EZF beamformer also conforms \textbf{Theorem}~\ref{theom2}, i.e. $\{{\bf{P}}^{\text{ Local EZF}}_{i,k}, k \in \mathcal{U}_i \}$ are in the column space of ${\bar{\bf{H}}}_{i}$.
Therefore, the local EZF beamformer also has its low-dimension substitution, which can be used to initialize the RWMMSE algorithm.
Specifically, the Local EZF beamformer ${\mathbf{P}}_i^{\text{Local EZF}}$ in \eqref{P_Local_EZF} can be rewritten as
\begin{equation}\label{P_Local_EZF_1}
      {\mathbf{P}}_i^{\text{Local EZF}} = {{{\bf{\tilde H}}}_i}^{H} {{{\bf{\bar X}}}_i^{\text{Local EZF}}} 
      \end{equation}
where $\tilde{{\bf{H}}}_i =  {\left[ {{\bf{H}}_{{i},k_1}^{H},{\bf{H}}_{{i},k_2}^{H}, \ldots ,{\bf{H}}_{i,k_{{\left| {{{\cal U}_i}} \right|}}}^{H}} \right]}^{H} \in \mathbb{C}^{\sum_{k \in \mathcal{U}_i} N_k \times M_i}$ denotes the channel matrix between AP $i$ and its serving UEs, and ${{{\bf{\tilde H}}}_i^H}$ can be expressed as
\begin{equation}
        {{{\bf{\tilde H}}}_i^H} \!= {\left[ {{{{\bf{\bar V}}}_{i,k_1}}{\bf{\bar \Sigma }}_{i,k_1}^H{\bf{\bar U}}_{i,k_1}^H, \ldots ,{{{\bf{\bar V}}}_{i,k_{\left| {{\mathcal{U}_i}} \right|}}}{\bf{\bar \Sigma }}_{i,k_{\left| {{\mathcal{U}_i}} \right|}}^H{\bf{\bar U}}_{i,k_{\left| {{\mathcal{U}_i}} \right|}}^H} \right]},
\end{equation}
and the block diagonal matrix $\bar{{\bf{X}}}_i^{\text{Local EZF}}$ 
is a low-dimension substitution of ${\mathbf{P}}_i^{\text{Local EZF}}$, given by
\begin{equation}
\begin{aligned}
\hspace{-3mm}
	  & {{{\bf{\bar X}}}_i^{\text{Local EZF}}}  \\
	   &\!\!  = \!\operatorname{blkdiag}\!\left(\!
\bar{\mathbf{U}}_{i, k_1}\!\!\left(\!\bar{\boldsymbol{\Sigma}}_{i, k_1}^{H}\!\right)\!\!^{-1},\dots,\! \bar{\mathbf{U}}_{i,k_{\left|\mathcal{U}_{i}\right|}}\!\!\left(\bar{\boldsymbol{\Sigma}}_{i,k_{\left|\mathcal{U}_{i}\right|}}^{H}\right)\!\!^{-1}\!\right)
{\bf{\Gamma}}_i
\end{aligned}
\end{equation}
with ${\bf{\Gamma}}_i = \frac{{{{\left( {{\bf{\bar V}}_i^H{{{\bf{\bar V}}}_i}} \right)}^{ - 1}}}}{{{{\left\| {{{\left( {{\bf{\bar V}}_i^H} \right)}^\dag }} \right\|}_\text{F}}}}\sqrt {{P_{\max ,i}}} $.

\begin{algorithm}[t]
    \renewcommand{\algorithmicrequire}{\textbf{Interact:}}
    \renewcommand{\algorithmicensure}{\textbf{Output:}}
    \caption{Fully Distributed Local EZF Algorithm}  \label{al:local_EZF}
    \begin{algorithmic}[1]
    \STATE At each AP $i$, \textbf{do} the following: \\
    \STATE Perform SVD on $\{{\bf{H}}_{i,k}, k\in \mathcal{U}_i \}$;\\
    \STATE Concat $\bar{\mathbf{V}}_i$ via~\eqref{V_i} as its effective channel;\\
    \STATE Calculate the Local EZF beamformer ${\mathbf{P}}_i^{\text{Local EZF}}$ via \eqref{P_Local_EZF}.
    \end{algorithmic}
\end{algorithm}
%
%
%
%
%
%
\begin{figure}[t]
    \centering
    \hspace{-6mm}
    \begin{minipage}[t]{0.45\columnwidth}
      \centering
      \subfigure[WMMSE]{
  \includegraphics[width=3.5cm]{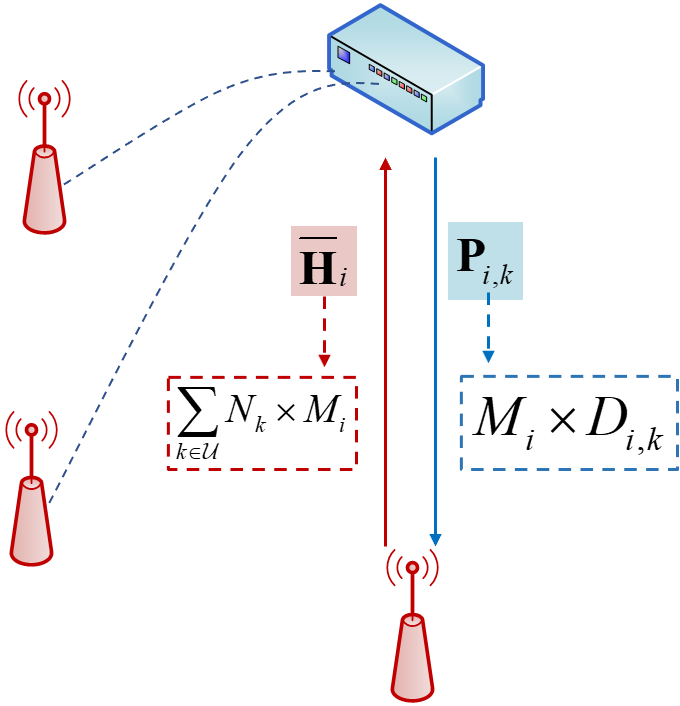}
        \label{Interaction in WMMSE }
      }
    \end{minipage}
    \begin{minipage}[t]{0.45\columnwidth}
      \centering
      \subfigure[RWMMSE]{
        \includegraphics[width=4cm]{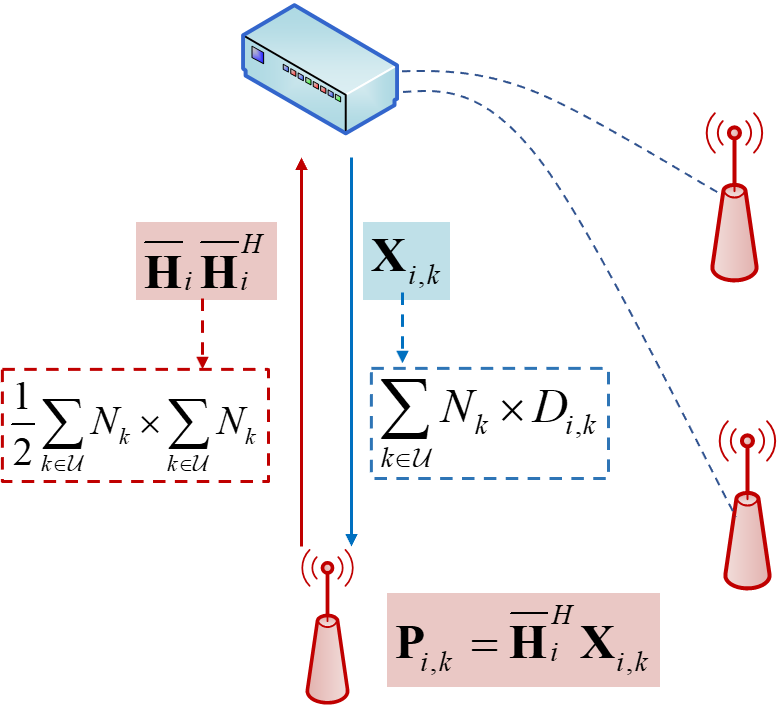}
        \label{Interaction in RWMMSE}
      }
    \end{minipage}
    \caption{An exemplary illustration of the interaction between the APs and the CUs.}
    \label{sdsjhf}
    \vspace{-8pt}
\end{figure}
\vspace{-10pt}
\subsection{Interaction and Computational Complexity Analysis}\label{subsec_EZF}
In this subsection, we analyze
 the interaction and computational complexity of the WMMSE algorithm, the low-interaction RWMMSE algorithm, and the Local EZF algorithm. 
As mentioned in \textbf{Remark}~\ref{remark1}, the current lowest computational complexity for WSR maximization in NCJT is $\mathcal{O}(M_i^3)$.
Considering the general status of modern communication networks, we assume that the number of transmit antennas on AP $i$ is greater than the total number of receive antennas, i.e., $M_i \geq \sum_{k \in \mathcal{U}} N_k$. 

\subsubsection{Interaction} In the WMMSE algorithm, each AP $i$ needs to transmit their channel matrices $\{ \bar{{\bf{H}}}_i  \in \mathbb{C}^{\sum_{k \in \mathcal{U}} N_k \times M_i}\}$ to their connected CUs, and CUs need to transmit the beamformers $\{{\bf{P}}_{i,k}^* \in \mathbb{C}^{M_i \times D_{i,k}}\}$ back to their controlling APs.
In the RWMMSE algorithm, the APs transmit $\{\bar{{\bf{H}}}_i\bar{{\bf{H}}}_{i}^H \in \mathbb{C}^{\sum_{k \in \mathcal{U}} N_k \times \sum_{k \in \mathcal{U}} N_k}\}$ to their connected CUs.
Since $\bar{{\bf{H}}}_i\bar{{\bf{H}}}_{i}^H$ is a symmetric matrix, only its upper triangular part needs to be transmitted. 
The CUs then transmit the low-dimension substitution $\{{\bf{X}}_{i,k}^*\in \mathbb{C}^{\sum_{k \in \mathcal{U}} N_k \times D_{i,k}}\}$ back to their controlling APs.
The Local EZF algorithm requires only local CSI, thereby eliminating the need for any interaction across networks.
In Fig.~\ref{sdsjhf}, we illustrate the interaction of the proposed WMMSE and RWMMSE algorithms.

\begin{table}[t]
\vspace{-8pt}
\renewcommand\arraystretch{1.5}
\setlength{\tabcolsep}{2 mm}{} 
  \centering
  
  \caption{Comparison of the interaction and complexity among the proposed algorithms}
    \vspace{-5bp}
  \label{tb.compar_alg_2}
  \resizebox{1\linewidth}{!}{
    \begin{tabular}{|c|c|c|}
      \hline \hline
      & \textbf{Interaction} 
& \textbf{ Complexity} \\
      \hline
      Local EZF
       & $0$
       & ${\cal O}\left( {{{\left( {\sum\limits_{k \in {{\cal U}_i}} {N_k}}  \right)}^2}{M_i}} \right)$ \\
      \hline
      WMMSE
       & {$ \left(\sum\limits_{k \in  \mathcal{U}}N_k+ {\sum\limits_{k \in  \mathcal{U}_i}D_{i,k}}\right)M_i $}
       & $\mathcal{O}\left({M_i}^{3}\right)$ \\
      \hline
      RWMMSE
       & $\left({\frac{1}{2}\sum\limits_{k \in \mathcal{U}}N_k } + \sum\limits_{k \in \mathcal{U}_i}D_{i,k} \right)  {\sum\limits_{k \in \mathcal{U}}N_k} $
       & $\mathcal{O}\left(  \left({\sum\limits_{k \in \mathcal{U}} N_k}\right)^2M_i \right)$ \\  
      \hline
      \hline
    \end{tabular}
  }
  \vspace{-5bp}
\end{table}

\subsubsection{Computational Complexity} 
Each iteration in the WMMSE algorithm and the RWMMSE algorithm calculates matrix inverse of $\left( {{{\bf{N}}_k} + {{\bf{H}}_k}{{\bf{P}}_k}{\bf{ P}}_k^H{\bf{H}}_k^H} \right) \in \mathbb{C}^{N_k \times N_k}$ and
${{\bf{E}}_k^*}\!\!\!\! \!\in\!\!\! \!\mathbb{C}^{D^k\! \times\! D^k}$ for $\mathbf{U}_k$ and $\mathbf{W}\!_k$ updates, respectively.
Besides, in order to solve $\{ {\bf{P}}_{i,k}^* \}$ (in the WMMSE algorithm) and $\{ {\bf{X}}_{i,k}^* \}$ (in the RWMMSE algorithm), matrices inverse of $\left( {\sum_{l \in \mathcal{U}}\alpha_l { {{\bf{H}}_{i,l}^H{{\bf{A}}_l}{{\bf{H}}_{i,l}}}  + {\mu _i}{\bf{I}}} } \right)$ and $\left( {\sum_{l \in \mathcal{U}} \alpha_l { {\bar{{\bf{H}}}_i{\bf{H}}_{i,l}^H{{\bf{A}}_l}{{\bf{H}}_{i,l}}\bar{{\bf{H}}}_i^H}  + {\lambda _i}{\bar{{\bf{H}}}_i\bar{{\bf{H}}}_i^H}} } \right)$ with dimensions ${M_i\times M_i}$ and $\sum_{k \in \mathcal{U}}\! N_k\!\! \times \!\!\sum_{k \in \mathcal{U}} N_k$  are required.
The Local EZF method needs to perform SVD on the channel matrix ${\bf{H}}_{i,k} \in \mathbb{C}^{N_k \times M_i}$, and calculate pseudo-inverse of ${\bar{\bV}_i^H} \in \mathbb{C}^{ \sum_{k \in \mathcal{U}_i} D_{i,k} \times M_i}$.
%

See Table \ref{tb.compar_alg_2} for a summary of both the interaction and the computational complexity of the three proposed algorithms.
We can see that the Local EZF method achieves the lowest interaction and computational complexity.
Both the interaction and the computational complexity of the RWMMSE algorithm is lower than its equivalent WMMSE algorithm.


\section{Joint Beamforming and Stream Allocation} \label{sec_SA}
In this section, we investigate the joint beamforming and stream allocation problem \eqref{WSR_problem} for WSR maximization in user-centric cell-free MIMO networks with NCJT. 
Due to the mix-integer and non-convexity nature, problem \eqref{WSR_problem} is NP-hard and thus is challenging to solve in general. 
By studying the beamforming structure with varying data streams, we propose a low-complexity joint beamforming and linear stream allocation algorithm, termed as RWMMSE-LSA.


%
\vspace{-5pt}
\subsection{Decoupling Beamforming Design and Stream Allocation}
For ease of illustration, we rewrite problem \eqref{WSR_problem} in a concise form using the definitions of $\mathbf{H}_k$ and $\mathbf{P}_k$ as follows
\begin{subequations} \label{Stream_limited_Problem_0}
\setlength{\abovedisplayskip}{1pt}
    \begin{align}
        \max_{\{\mathbf{P}_{i,k}\},\{D_{i,k}\}} \  \  & \sum\limits_{k\in\mathcal{U}} {\log \det \left( {\bf{I}} + {\bf{H}}_k {\bf{P}}_k {{\bf{P}}_k^H} {{\bf{H}}_k^H} {{\mathbf{N}}_k^{ - 1}} \right)}  \\
        \operatorname{ s.t. } \qquad  \      &  \eqref{cons_p} , \ \eqref{stream_a}. 
    \end{align}
    \setlength{\belowdisplayskip}{0.5pt}
\end{subequations}
The main difficulty of solving problem \eqref{Stream_limited_Problem_0} roots in the coupled decision variables, i.e., the number of streams $D_{i,k}$ and the beamformer $\mathbf{P}_{i,k} \in \mathbb{C}^{M_i \times D_{i,k}}$.
In particular, $D_{i,k}$ determines the size of the beamformer $\{\mathbf{P}_{i,k}\}$.

In order to decouple the decision variables $\{\mathbf{P}_{i,k}\}$ and $\{D_{i,k}\}$, we introduce a binary diagonal stream indicator matrix ${\bf{L}}_{i,k} \in \mathbb{B}^{N_k \times N_k}$ for transmit stream allocation and a virtual beamformer ${\bar{\bf{P}}_{i,k}}\in \mathbb{C}^{M_i \times N_k}$ with fixed dimension assuming UE $k$ receives the maximum number of streams $N_k$ from AP~$i$. 
The actual total number of streams that AP $i$ transmits to UE~$k$ is $\operatorname{Tr}\left( {\bf{L}}_{i,k}\right)=D_{i,k}$. The columns in ${\bar{\bf{P}}_{i,k}}$ that are selected as the actual beamformer ${\bf{P}}_{i,k}$ (where the corresponding diagonal elements of ${\bf{L}}_{i,k}$ are one) can be represented by ${\bar{\bf{P}}_{i,k}}{\bf{L}}_{i,k}$.


Replacing ${\bf{P}}_{i,k}$ with ${\bar{\bf{P}}_{i,k}}{\bf{L}}_{i,k}$, the achievable data rate of UE $k$ in \eqref{R_k} can be rewritten as 
\begin{equation}\label{R_k_LSA}
\setlength{\abovedisplayskip}{1pt}
{\log \det \left( {{\bf{I}}\! +\! \left(\sum\limits_{i \in {{\cal I}_k}} {{\bf{H}}_{i,k}^{}{\bar{\bf{P}}_{i,k}}{\bf{L}}_{i,k}{\bf{L}}_{i,k}^H\bar{\bf{P}}_{i,k}^H{\bf{H}}_{i,k}^H}\right)  \bar{\mathbf{N}}_k^{ - 1}} \right)}
\setlength{\belowdisplayskip}{0.5pt}
\end{equation}
where $	\bar{\bf{N}}_k =  \sum_{l \in {\cal U}_{-k}} {\sum_{j \in {{\cal I}_l}} {{\bf{H}}_{j,k}{\bar{\bf{P}}_{j,l}}}\mathbf{L}_{j,l}\mathbf{L}_{j,l}^H {\bar{\bf{P}}_{j,l}^H} {\bf{H}}_{j,k}^H }+ \sigma _k^2{\bf{I}}$.

Problem \eqref{Stream_limited_Problem_0} can then be equivalently transformed to 
\begin{subequations}\label{Stream_limited_Problem_1}
    \begin{align}
        \underset{\substack{\{\bar{\mathbf{P}}_{i,k}\}, \\ \{\mathbf{L}_{i,k}\}}}{\max}  \  \  & \sum\limits_{k\in\mathcal{U}}\!\alpha_k {\log \det \left( {\bf{I}} + {\bf{H}}_k \bar{\bf{P}}_k {\bf{L}}_k {\bf{L}}_k {\bar{\bf{P}}_k^H} {{\bf{H}}_k^H} {\bar{\mathbf{N}}_k^{ - 1}} \right)} \\        \operatorname{ s.t. } \quad  \      &  \eqref{cons_p} ,\\
        & \sum_{i \in \mathcal{I}_k}{\Tr{\left( {\bf{L}}_{i,k} \right)}} \le N_k, \ \forall k,\label{stream_c}\\
        & \ {\bf{L}}_{i,k} = \operatorname{diag}\left(l_{i,k}^{(1)},\dots, l_{i,k}^{(N_k)} \right), \ \forall k, \ \forall i,\label{stream_d1} \\
        & \  l_{i,k}^{(m)}  \in \{0,1 \}, \ \forall k,\ \forall i, \ m=1,\dots,N_k\label{stream_d2}
    \end{align}
\end{subequations}
where ${\bar{\bf{P}}_k}\! \triangleq \!{\rm{blkdiag}}\left( {\bar{\bf{P}}_{{i_1},k}, \ldots ,\bar{\bf{P}}_{{i_{\left| {{{\cal I}_k}} \right|}},k}^{}} \right)\in \mathbb{C}^{M^k \times {|\mathcal{I}_k|N_k}}$, 
and  ${{\bf{L}}_{k}}\! \triangleq\operatorname{blkdiag}\left( {\bf{L}}_{i_1,k},\! \dots,\!{\bf{L}}_{i_{|\mathcal{I}_k|},k} \!\right)\!\in \!\mathbb{B}^{{|\mathcal{I}_k|N_k} \times {|\mathcal{I}_k|N_k}}$.
We  see that the original optimization problem \eqref{Stream_limited_Problem_0} with coupled variables $\{\mathbf{P}_{i,k}\}$ and $\{D_{i,k}\}$ is now equivalently transformed to problem \eqref{Stream_limited_Problem_1} with independent variables $\{{\bar{\bf{P}}_{i,k}}\}$ and $\{{\bf{L}}_{i,k}\}$.

\subsection{WMMSE Transformation and Problem Reformulation} 
It is well known that mixed-integer and nonconvex problems such as \eqref{Stream_limited_Problem_1} are NP-hard and finding their global optimum is generally difficult.
	The classic branch-and-bound method has exponential  complexity \cite{bertsekas1997nonlinear}.
	In \cite{9500429,8482453}, each integer variable is rewritten as an $\ell_{0}$-norm and then approximated by a weighted $\ell_{1}$-norm to relax the original integer variables to continuous variables.
	However,
the initialization of the above algorithms in each iteration can be difficult due to the coupled variables, while we have decoupled them in problem \eqref{Stream_limited_Problem_1}.
%

Due to the above discrepancies, we first utilize the WMMSE transformation to reformulate problem \eqref{Stream_limited_Problem_1}, and consider to leverage the low-dimension substitution property together with the BCD approach, to provide low-complexity beamforming and stream allocation solutions.
\begin{subequations}\label{Stream_limited_Problem}
    \begin{align}
    \hspace{-2mm}
       \underset{\substack{{\{\!\bf{U}}_k\!\},\{{\bf{W}}_k\succ {\bf{0}}\}, \\ \left\{ {{\bar{\bf{P}}\!_{i,k}}} \right\} ,\{{\bf{L}}_{i,k}\}}}{\max}   & \sum\limits_{k\in\mathcal{U}} \!\alpha_k\!\! \left( {\log \det \left( {{{\bf{W}}_k}} \right) \!\!- \!\!{\rm{Tr}}\!\left( {{{\bf{W}}\!_k}{{\bf{E}}_k}\!\!\left( {\bf{U}}_k,\!\bar{\bf{P}} \right)}\! \right)}\!\right) \\
        \operatorname{ s.t. } \qquad      &  \sum\limits_{k\in \mathcal{U}_i}\!\!\operatorname{Tr}\left( \bar{\mathbf{P}}_{i,k}\mathbf{L}_{i,k}\mathbf{L}_{i,k}^H \bar{\mathbf{P}}_{i,k}^{H}\right)\leq P_{\max,i}, \ \forall i,\\
        & \eqref{stream_c}, \ \eqref{stream_d1}, \ \eqref{stream_d2}.
    \end{align}
\end{subequations}
Considering the presence of the integer variables $\{{\mathbf{L}}_{i,k}\}$, directly using the classic BCD approach to problem \eqref{Stream_limited_Problem} is basically challenging.
An intuitive approach is to relax the integer constraints \eqref{stream_d2} into continuous constraints, i.e., $l_{i,k}^{(m)}  \in [0, \ 1]$. 
However, this approach cannot guarantee that the relaxed optimization problem converges to a solution that yields the original integer constraints \eqref{stream_d2}.
In contrast, we propose a linear stream allocation approach to problem~\eqref{Stream_limited_Problem}, based on the following unique observation on the identity of the stream indicator matrix ${\bf{L}}_{i,k}$.
	
\begin{remark}\label{prop_L} (Quadratic-Linear Property):
	Since $\{\mathbf{L}_{i,k}\}$ are diagonal matrices with values of zero or one on the diagonal, we have:
 \begin{equation}\label{L_identity}
 	 {\bf{L}}_{i,k}{\bf{L}}_{i,k}^H = {\bf{L}}_{i,k}, \ \forall k, \ \forall i.
 	 \setlength{\belowdisplayskip}{5pt}
 \end{equation}
 \setlength{\belowdisplayskip}{5pt}
\end{remark}
Combining \textbf{Remark}~\ref{prop_L} and relaxing the 0-1 integer constraints~\eqref{stream_d2} to the continuous constraints \eqref{stream_d3}, problem \eqref{Stream_limited_Problem} can be relaxed to 
\begin{subequations}
\label{Stream_limited_Problem_c}
\begin{align}
\hspace{-2mm}
       \underset{\substack{{\{\bf{U}}_k\!\},\{{\bf{W}}_k\succ {\bf{0}}\}, \\ \left\{ {{\bar{\bf{P}}\!_{i,k}}} \right\} ,\{{\bf{L}}_{i,k}\}}}{\max}   & \sum\limits_{k\in\mathcal{U}}\alpha_k \left(\log \det \left( {{{\bf{W}}_k}} \right)
       +2{\rm{Tr}}\left(\mathbf{W}_k\mathbf{U}_k^H\mathbf{H}_k\bar{\mathbf{P}}_k \right)\right)\notag \\
       &\! -\!\!\! \sum\limits_{k\in\mathcal{U}} \alpha_k{\rm{Tr}}\!\left( {{\bf{W}}\!_k}\mathbf{U}_k^H\mathbf{H}_k\bar{\mathbf{P}}_k\mathbf{L}_k\bar{\mathbf{P}}_k^H\mathbf{H}_k^H\mathbf{U}_k \right)  \\
       &-\sum\limits_{k\in\mathcal{U}} \alpha_k {\rm{Tr}} \left( \mathbf{W}_k\mathbf{U}_k^H\bar{\mathbf{N}}_k\mathbf{U}_k \right) \notag \\
        \operatorname{ s.t. } \qquad      &  \sum\limits_{k\in \mathcal{U}_i}\!\!\operatorname{Tr}\!\left( \bar{\mathbf{P}}_{i,k}\mathbf{L}_{i,k}\bar{\mathbf{P}}_{i,k}^{H}\right)\leq P_{\max,i}, \ \forall i,\label{cons_L_linear} \\
        & \eqref{stream_c}, \ \eqref{stream_d1}, \ \eqref{stream_d2}.
        \\
        & l_{i,k}^{(m)}  \in [0,1 ], \ \forall k,\ \forall i,\ m=1,\dots,N_k. \label{stream_d3}
\end{align}
\end{subequations}
Note that problem \eqref{Stream_limited_Problem_c} is linear w.r.t the decision variables $\{\mathbf{L}_{i,k}\}$, and is convex w.r.t. the decision variables $\{{\bf{W}}_k\}$, $\{{\bf{U}}_k\}$ and $\{\bar{\mathbf{P}}_{i,k}\}$.
%
%
%
\subsection{RWMMSE-LSA Algorithm}
In the following, we use the low-dimension substitution property in \textbf{Theorem}~\ref{theom2} together with the BCD method to provide efficient solutions to problem \eqref{Stream_limited_Problem_c}. 
Using results in \textbf{Theorem}~\ref{theom2}, we can rewrite $\bar{\mathbf{P}}_{i,k}$ as
\begin{equation}\label{P_X_bar}
	\bar{\mathbf{P}}_{i,k} = \bar{\mathbf{H}}_i^H\bar{\mathbf{X}}_{i,k}
\end{equation} 
where $\bar{\bf{X}}_{i,k} \in {\mathbb{C}}^{{\sum_{k\in \mathcal{U}}N_k}\times N_k}$ is a low-dimension substitution of $\bar{\mathbf{P}}_{i,k}$. 
We define 
\begin{equation}\label{hat_X}
	\tilde{\bf{X}}_{i,k} \triangleq {\bar{\bf{X}}_{i,k}}{\bf{L}}_{i,k}\in {\mathbb{C}}^{{\sum_{k\in \mathcal{U}}N_k}\times N_k}
\end{equation} 
as the actual low-dimension beamformer substitution (non-zero columns) that carries the allocated data streams from AP~$i$ to UE $k$. 

Following the BCD approach, we first fix $\{\bar{\bf{X}}_{i,k}\}$, $\{{\bf{L}}_{i,k}\}$ and update $\{{\bf{W}}_k^*\}$, $\{{\bf{U}}_k^*\}$ by substituting \eqref{hat_X} into \eqref{MU_U_MMSE_rwmmse} and \eqref{MU_W_rwmmse}. 
%
%
%
With fixed $\{{\bf{W}}_k^*\}$ and $\{{\bf{U}}_k^*\}$, we update ${\bar{\bf{X}}_{i,k}^*}$ by
\begin{equation}
\label{RWMMSE_X_stream}
\begin{aligned}
        {\tilde{\bf{X}}}_{i,k}^*  = & {\left( {\sum\limits_{l \in \mathcal{U}}\alpha_l { {\bar{{\bf{H}}}_i{\bf{H}}_{i,l}^H{{\bf{A}}_l}{{\bf{H}}_{i,l}}\bar{{\bf{H}}}_i^H}  + {\lambda _i}{\bar{{\bf{H}}}_i\bar{{\bf{H}}}_i^H}} } \right)^{ - 1}}\\
        &\times \alpha_k\bar{{\bf{H}}}_i{\bf{H}}_{i,k}^H{\bf{U}}_k^{*}{\bf{W}}_k^*{\bf{\Xi }}_{i,k}^H{\bf{L}}_{i,k}.
\end{aligned}
\end{equation}

Then fixing $\{{\bf{W}}_k^*\}$, $\{{\bf{U}}_k^*\}$ and $\{\bar{\mathbf{X}}_{i,k}^*
\}$ (resp. $\{\bar{\mathbf{P}}_{i,k}^*
\}$ in \eqref{P_X_bar}), the linear stream allocation problem can be expressed as
%
\begin{equation}
\label{RWMMSE_Probelm_L_i_k_stream_limited}
\hspace{-2mm}
    \begin{aligned}
\min_{\{{\bf{L}}_{i,k}\}}
&\sum\limits_{k\in\mathcal{U}}\sum\limits_{i \in \mathcal{I}_k} \alpha_k {\rm{Tr}}\left({(\bar{ {{\bf{P}}}}_{i,k}^*)^H{\bf{H}}_{i,k}^H{{\bf{A}}_k}{{\bf{H}}_{i,k}}{\bar{{\bf{P}}}_{i,k}^*}{\bf{L}}_{i,k}} \right)   \\
&\!\! -\!\! 2\!\sum\limits_{k\in\mathcal{U}} \! \sum\limits_{i \in \mathcal{I}_k}\!\!\alpha_k{\mathop{\rm Re}\nolimits}\! \left\{\! {{\rm{Tr}}\!\!\left( {{{\bf{\Xi }}_{i,k}}\!{{\bf{W}}_k^*}({\bf{U}}_k^*)\!^H\!{{\bf{H}}_{i,k}}{\bar{{\bf{P}}}_{i,k}^*}}{\bf{L}}_{i,k} \right)}\right\}  \\
& +\!\! \sum\limits_{k\in\mathcal{U}} \!\! \alpha_k {{\rm{Tr}}\!\!\left(\! { 
\sum\limits_{j \in \mathcal{I}_l} \! \sum\limits_{l \in {\cal U}_{-k}} \!\!({\bar{\bf{P}}}_{j,l}^*)\!^H\!{\bf{H}}_{j,k}^H{{\bf{A}}_k}{{{\bf{H}}_{j,k}}{{\bar{\bf{P}}}_{j,l}^*}{\bf{L}}_{j,l}}  }\!\! \right)}  \\
\operatorname{ s.t. } \ & \eqref{stream_c}, \ \eqref{stream_d1}, \ \eqref{cons_L_linear}, \ \eqref{stream_d3}. \\
    \end{aligned}
\end{equation}
By extracting ${{\bf{L}}_{k}}$ related items from \eqref{RWMMSE_Probelm_L_i_k_stream_limited}, the stream allocation problem for each UE $k$ can be written as
\begin{equation}
\label{RWMMSE_Probelm_L_i_k_stream_limited_k}
    \begin{aligned}
\min_{{\bf{L}}_{k}} \
&\sum\limits_{i \in \mathcal{I}_k} \sum\limits_{l \in {\cal U}} \alpha_l {\rm{Tr}}\left( {(\bar{{\bf{P}}}_{i,k}^*)^H{\bf{H}}_{i,l}^H{{\bf{A}}_l}{{\bf{H}}_{i,l}}{\bar{{\bf{P}}}_{i,k}^*}{\bf{L}}_{i,k}} \right)  \\
&\!\! - \!\!2\!\!\sum\limits_{i \in \mathcal{I}_k}\!\!\alpha_k{\mathop{\rm Re}\nolimits} \left\{ {{\rm{Tr}}\left( {{{\bf{\Xi }}_{i,k}}{{\bf{W}}_k^*}({\bf{U}}_k^*)^H{{\bf{H}}_{i,k}}{\bar{{\bf{P}}}_{i,k}^*}}{\bf{L}}_{i,k} \right)} \right\}  \\
\operatorname{ s.t. } \ & \eqref{stream_c}, \ \eqref{stream_d1}, \ \eqref{cons_L_linear}, \ \eqref{stream_d3}.
\end{aligned}
\end{equation}
\begin{lemma}
	\label{Prop_cons_remove}
The individual power constraints \eqref{cons_L_linear} in the stream allocation problem \eqref{RWMMSE_Probelm_L_i_k_stream_limited_k} are always fulfilled.
\end{lemma}
\begin{IEEEproof}
See Appendix \ref{Appendix_Prop_cons_remove}.
\end{IEEEproof}
%
%
According to \textbf{Lemma}~\ref{Prop_cons_remove}, constraints (42b) can be excluded from \eqref{RWMMSE_Probelm_L_i_k_stream_limited_k}. Thus, we can easily obtain a closed-form optimal solution to problem \eqref{RWMMSE_Probelm_L_i_k_stream_limited_k}, which readily satisfies the 0-1 integer constraints \eqref{stream_d2}.
%
The derivation is as follows.
Problem \eqref{RWMMSE_Probelm_L_i_k_stream_limited_k} can be equivalently rewritten as
\begin{equation}\label{Pro_L_k_simple}
	\begin{aligned}
\min_{{\bf{L}}_{k}} \
&{\rm{Tr}}( {{\bf{\Psi}}_{k}}{\bf{L}}_{k})  \\
\operatorname{ s.t. } \ & {{\bf{L}}_{k}}(l) \in [0,1],\ \forall  l = 1,\dots, {|\mathcal{I}_k|N_k}, \\
&{\rm{Tr}}( {\bf{L}}_{k}) \leq N_k
\end{aligned}
\end{equation}
where ${{\bf{L}}_{k}}(l)$ denotes the $l$th diagonal element of ${{\bf{L}}_{k}}$, $	{{\bf{\Psi}}_{k}} =  \operatorname{blkdiag}\left( \operatorname{diag}( {\bf{\Psi}}_{i_1,k}),\dots, \operatorname{diag}({\bf{\Psi}}_{i_{|\mathcal{I}_k|},k} )\right)$ 
with ${\bf{\Psi}}_{i,k}  \triangleq 	\sum\limits_{l \in {\cal U}}\!\! \alpha_l {(\!\bar{{\bf{P}}}_{i,k}^*\!)\!^H\!{\bf{H}}_{i,l}^H{{\bf{A}}_l}{{\bf{H}}_{i,l}}{\bar{{\bf{P}}}_{i,k}^*}}\!\!-
 2\alpha_k{\mathop{\rm Re}\nolimits}\!\! \left\{\! { {{{\bf{\Xi }}_{i,k}}\!{{\bf{W}}_k^*}({\bf{U}}_k^*)\!^H\!{{\bf{H}}_{i,k}}{\bar{{\bf{P}}}_{i,k}^*}}}\! \right\}$.
The optimal solution to problem \eqref{Pro_L_k_simple} is  selecting the smallest $\pi_k \leq N_k $ diagonal elements of ${{\bf{\Psi}}_{k}}$ that are less than zero, i.e.,
the optimal ${{\bf{L}}_{k}^*}$ to problem \eqref{RWMMSE_Probelm_L_i_k_stream_limited_k} is given by
\begin{equation} \label{LSA_L}
		{{\bf{L}}_{k}^*}(l)  =
\left\{\begin{aligned} 
& 1, \text{for} \ l \in \mathbf{\Pi}_k,\\
& 0, \text{o.w.}
\end{aligned}\right.
\end{equation}
where 
$\mathbf{\Pi}_k$ denotes the indexes of the $\pi_k$ smallest diagonal elements of ${{\bf{\Psi}}_{k}}$ that  ${{\bf{\Psi}}_{k}}(l)<0$.

In \textbf{Algorithm}~\ref{al:RWMMSE_LSA_MU}, we summarize the proposed RWMMSE-LSA algorithm.
Different from the complicated initialization process in \cite{8482453},
the stream indicator matrix $\mathbf{L}_{i,k}$ in our RWMMSE-LSA process can be randomly initialized to meet the maximum number of receive streams constraints \eqref{stream_a}.
After initializing $\mathbf{L}_{i,k}$, we can use our proposed Local EZF method in Section \ref{subsec_EZF} to initialize   $\bar{\mathbf{X}}_{i,k}$.
\begin{Proposition}\label{Prop_convergence}
(Convergence):
Any limit point of the iterative sequence generated by the RWMMSE-LSA algorithm is a stationary point of problem \eqref{WSR_problem}.
\end{Proposition} 
\begin{IEEEproof}
See Appendix~\ref{Appendix_Prop_convergence}.
\end{IEEEproof}

\begin{algorithm}[t]
    \renewcommand{\algorithmicrequire}{\textbf{Interact:}}
    \renewcommand{\algorithmicensure}{\textbf{Output:}}
    \caption{Joint Beamforming and Stream Allocation Algorithm (RWMMSE-LSA)}  \label{al:RWMMSE_LSA_MU}
    \begin{algorithmic}[1]
    \REQUIRE Each AP $i$ sends $\{\bar{{\bf{H}}}_i\bar{{\bf{H}}}_i^H\}$ to the connected CUs;\\
    \STATE Initialize  $\{{\bf{L}}_{i,k}\}$ and $\{\bar{\bf{X}}_{i,k}\}$ in CUs;\\
    \STATE \textbf{Repeat}~$\leftarrow$[CUs]\\
    \STATE  \quad Update $\{{\bf{U}}_k^*\}$ by substituting \eqref{hat_X} into \eqref{MU_U_MMSE_rwmmse};
    \STATE  \quad Update $\{{\bf{W}}_k^*\}$ by substituting \eqref{hat_X} into \eqref{MU_W_rwmmse};\\
    \STATE  \quad Update $\{\tilde{\bf{X}}_{i,k}^*\}$ via \eqref{RWMMSE_X_stream};\\
    \STATE  \quad Update $\{{\bf{L}}_{i,k}^*\}$ via \eqref{LSA_L};\\
    \STATE  \textbf{Until}~Converge\\
    \REQUIRE  CUs transmit $\{{\bf{X}}_{i,k}^*, k \! \in \! \mathcal{U}_i \}$ to the connected AP $i$;\\
    \ENSURE Each AP $i$ calculates $\{\mathbf{P}_{i,k}^*=\bar{\mathbf{H}}_{i}^H\mathbf{X}_{i,k}^*, k \in \mathcal{U}_i \}$.
    \end{algorithmic}
\end{algorithm}

\begin{remark} \label{remark_US} (User Scheduling):
	The proposed RWMMSE-LSA algorithm for stream allocation can be further extended to enable linear user scheduling.
	Specifically, if we set $\mathcal{I}_k = \mathcal{I}$, $\forall k$, the RWMMSE-LSA algorithm selects the serving APs to each UE $k$ among all APs through stream allocation to maximize the WSR.
	The APs corresponding to the selected streams are then chosen as the serving APs $\mathcal{I}_k$ for UE $k$, which we call the RWMMSE-LUS algorithm.
\end{remark}
\vspace{-10pt}
\section{Simulation Results} \label{sec_simu}
In this section, we present simulation results to illustrate the efficiency of the proposed algorithms. 
\vspace{-10pt}
\subsection{Simulation Setup}
We model the channel as Rayleigh channel with circularly symmetric standard complex normal distribution.
We model the pathloss as $128.1+37.6 \log _{10}(d)[\mathrm{dB}] $ \cite{dahrouj2010coordinated}, where $d \in [ 0.1 , 0.3]$ in kilometers denotes the distance between the AP and the UE. 
Without loss of generality, all APs (and UEs) are assumed to have the same number of antennas, i.e., $M_i = M$ and $N_k = N$. 
The transmit power budget is set to be $P_{\max}$ for all APs, the fairness weight $\alpha_k$ and noise power $\sigma_{k}^{2}=10^{\frac{1}{K} \sum_{k} \log _{10} \frac{1}{N_{k}M_i}\left\|\mathbf{H}_{k}\right\|_{F}^{2}} \times 10^{-\frac{\mathrm{SNR}}{10}}$ are set equally for all UEs, where SNR is the average received SNR for all UEs without beamforming. 
In addition, the size of the serving AP set for each UE $k$ is assumed to be same, i.e., $\left|\mathcal{I}_{k}\right|=L$, and the $L$ nearest APs are selected to serve UE~$k$. 
Monte Carlo simulations are performed over 100 randomly generated channel realizations. 
The zero-interaction Local EZF beamformer serves as the initial points of the WMMSE and the RWMMSE algorithms. 
Unless otherwise stated, the system parameters are summarized in Table \ref{tb.simu}.
\begin{table}
\vspace{-10bp}
\renewcommand\arraystretch{2}
\setlength{\tabcolsep}{1.5 mm}{} 
  \centering
  \caption{Simulation parameters}
  \vspace{-5bp}
  \label{tb.simu}
  \resizebox{0.9\columnwidth}{!}{
    \begin{tabular}{|c|c|c|}
      \hline \hline
      \makebox[0.3\columnwidth][c] & \makebox[0.3\columnwidth][c]{\textbf{Parameter}}
       & \makebox[0.3\columnwidth][c]{\textbf{Value}} \\
\hline
Network config. & $I,K,M,N$ & $4,8,64,4$\\
\hline
Power budget &$P_{\max}$ & $1 \text{W}$\\ 
\hline
Fairness weight & $\alpha_k$ & $1$\\
\hline
Size of serving set & $L$ & $2$\\
\hline 
\hline
    \end{tabular}
  }
\end{table}
\begin{figure}[t]
\centering
\includegraphics[width=0.68\columnwidth]{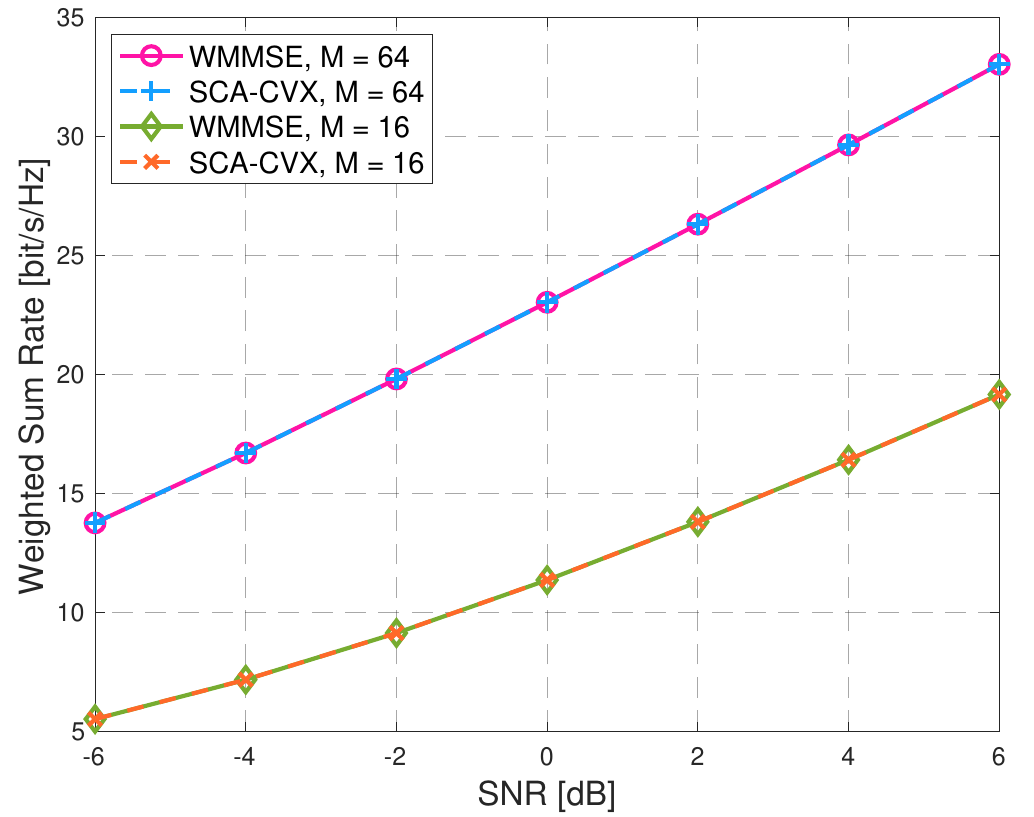}
\vspace{-8bp}
\caption{Comparison of WSR between the SCA-CVX algorithm in \cite{8482453} and the WMMSE baseline.}
\vspace{-8bp}
\label{Compare_SCA_WMMSE}
\end{figure}
\begin{figure}[!t]
\vspace{-4bp}
\centering
\includegraphics[width=0.68\columnwidth]{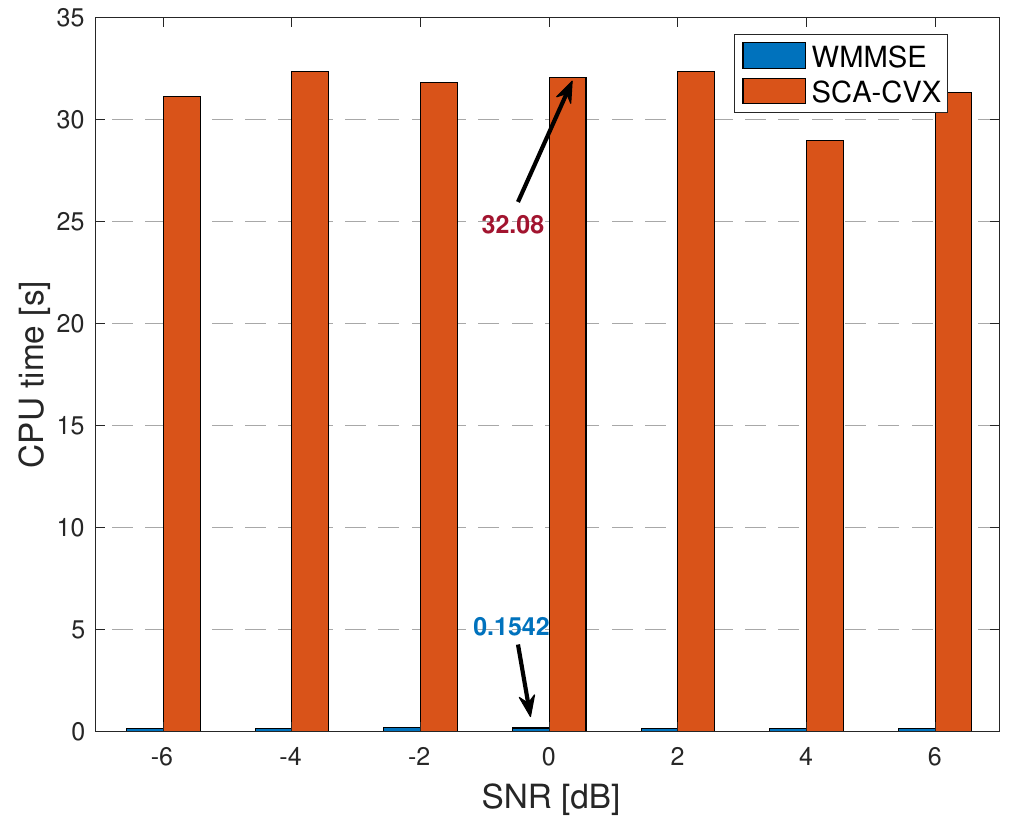}
\vspace{-8bp}
\caption{Comparison of average CPU time among FP based algorithm in \cite{ammar2021downlink}, SCA-CVX algorithm in \cite{8482453}  and WMMSE baseline.}
\label{Compare_SCA_WMMSE_CPU}
\vspace{-5bp}
\end{figure}
\vspace{-10pt}
\subsection{Comparison with the SCA Based Approach}
In Fig.~\ref{Compare_SCA_WMMSE} and Fig.~\ref{Compare_SCA_WMMSE_CPU}, we compare the WSR performance and the average CPU time between the WMMSE algorithm and the SCA-CVX algorithm in \cite{8482453} under different SNR values.
We set the number of receive antennas as $N = 1$ and the number of data streams as $D = 1$, since the SCA-CVX method is proposed for the single receive antenna case.
We can see that the SCA-CVX algorithm in \cite{8482453} and the WMMSE algorithm achieve the same WSR, but the SCA-CVX algorithm consumes 100x+ time than the WMMSE algorithm to converge. 
This behavior conforms to the complexity analysis in \textbf{Remark}~\ref{remark1} that the WMMSE approach has substantially lower complexity than the SCA based algorithms in \cite{8482453,vu2020noncoherent}.
In the following, we choose the WMMSE algorithm  as the baseline.
%
%
\vspace{-27pt}
\subsection{Comparison between WMMSE and RWMMSE}
\begin{figure}[!t]
\vspace{-8bp}
\centering
\includegraphics[width=0.68\columnwidth]{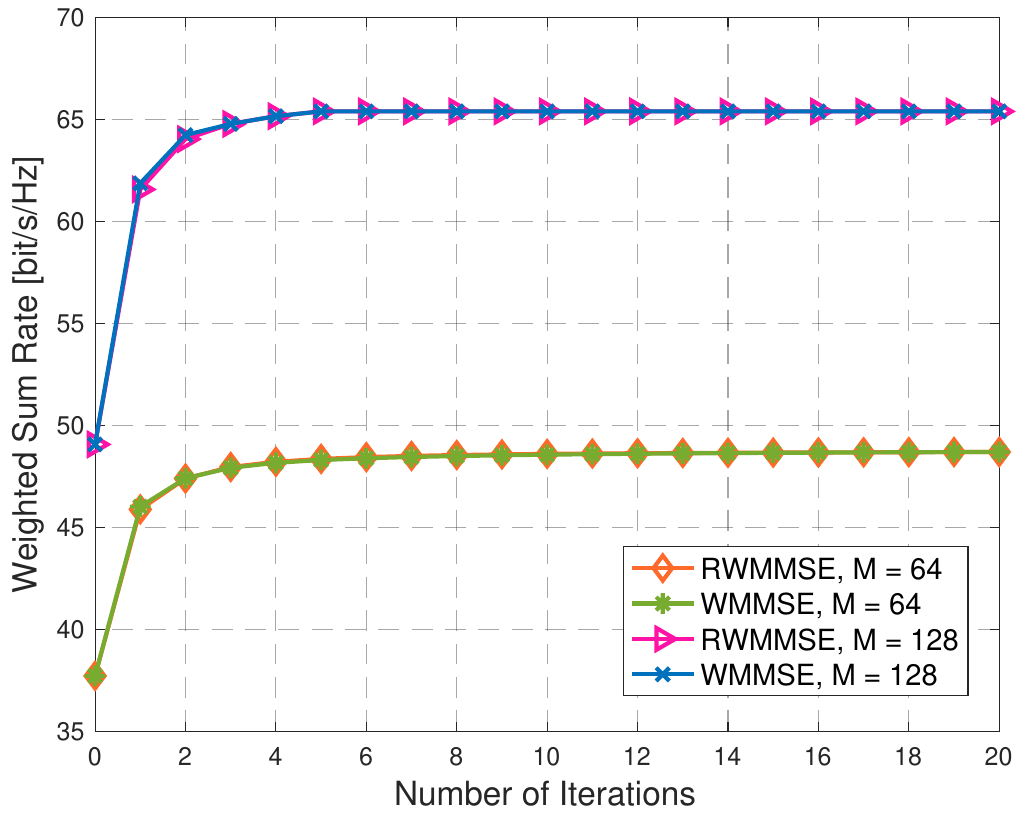}
\vspace{-10bp}
\caption{Convergence performance of the WMMSE and the RWMMSE algorithms, where SNR = $0$ [dB].}
\vspace{-8bp}
\label{Convergence_SNR}
\end{figure}

\begin{figure}[!t]
\centering
\includegraphics[width=0.68\columnwidth]{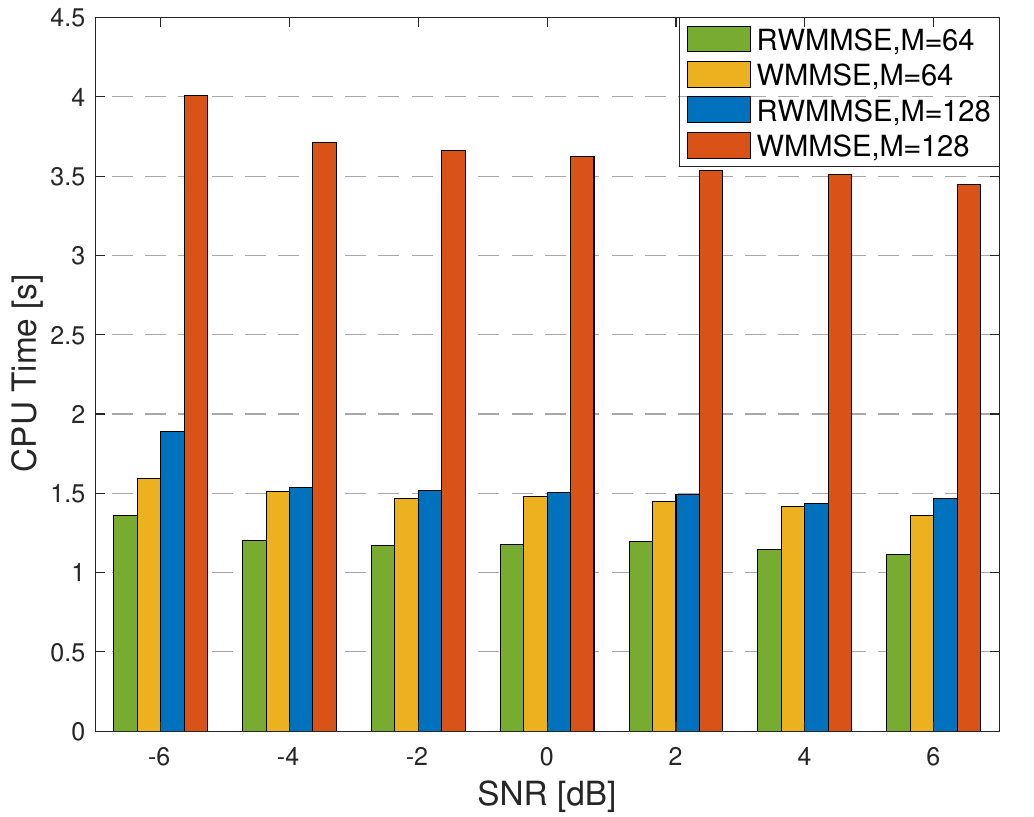}
\vspace{-8bp}
\caption{Comparison of average CPU time of the WMMSE algorithm and the RWMMSE algorithm.}
\label{CPU_RWMMSE_WMMSE}
\end{figure}

Fig. \ref{Convergence_SNR} depicts the convergence behavior of our proposed RWMMSE algorithm and the baseline WMMSE algorithm for the case of $\text{SNR} = 0$ [dB] with different numbers of transmit antennas.
We observe that the low-interaction RWMMSE and the WMMSE algorithms converge smoothly to the same WSR, which is consistent with \textbf{Theorem}~\ref{theom2}.
Moreover, Fig.~\ref{CPU_RWMMSE_WMMSE} shows that 
the baseline WMMSE algorithm generally requires more CPU time than the RWMMSE algorithm, particularly as the number of transmit antennas increases. 
The reason is that the WMMSE and the RWMMSE algorithms have linear and cubic complexity in $M$, respectively.
%
These observed phenomena demonstrate that our proposed RWMMSE algorithm can achieve the same WSR performance as the WMMSE algorithm, but with significantly lower computational complexity. 

\begin{figure}[!t]
\vspace{-12bp}
\centering
\includegraphics[width=0.68\columnwidth]{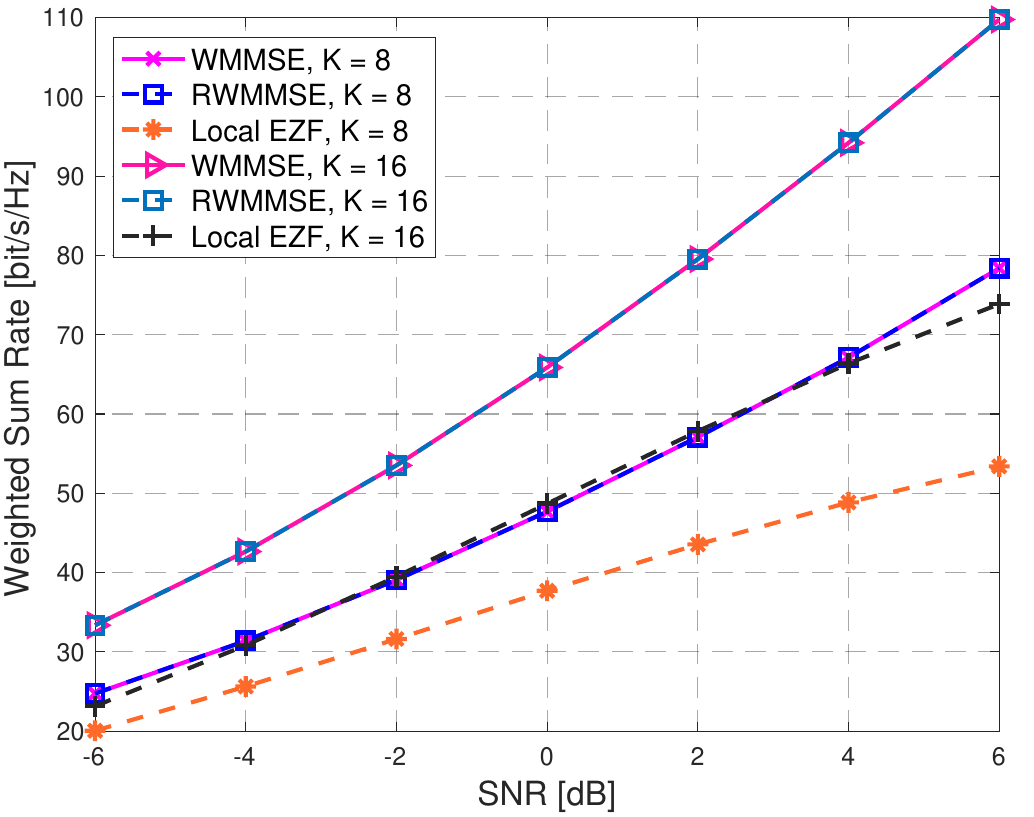}
\vspace{-10bp}
\caption{Comparison of WSR between the fully distributed Local EZF method, the baseline WMMSE algorithm and the  RWMMSE algorithm.}
\label{Comparison_Algorithms}
\vspace{-8pt}
\end{figure}
\vspace{-7pt}
\subsection{Comparison of the Proposed Beamforming Algorithms}
In Fig. \ref{Comparison_Algorithms}, we compare the WSR of the RWMMSE algorithm and the fully distributed Local EZF algorithm with the baseline WMMSE algorithm, under different number of SNR values and UEs.
We can see that the WMMSE algorithm and the RWMMSE algorithm yield almost the same WSR, both significantly outperforming the Local EZF method. 
In particular, when $K=16$ and SNR = $6$ [dB],  the fully distributed Local EZF method only achieves $73.8$ [bit/s/Hz], whereas the WMMSE/RWMMSE algorithm 
obtains a significantly higher WSR of 108.9 [bit/s/Hz], representing a 48$\%$ improvement over the Local EZF method.
The reason for the lower WSR achieved by the Local EZF method is that it only suppresses inter-user interference but does not optimize transmit power for WSR maximization.
%
%
%
\vspace{-10pt}
\subsection{Comparison between NCJT and CJT}
\begin{figure}[!t]
\vspace{-5bp}
\centering
\includegraphics[width=0.68\columnwidth]{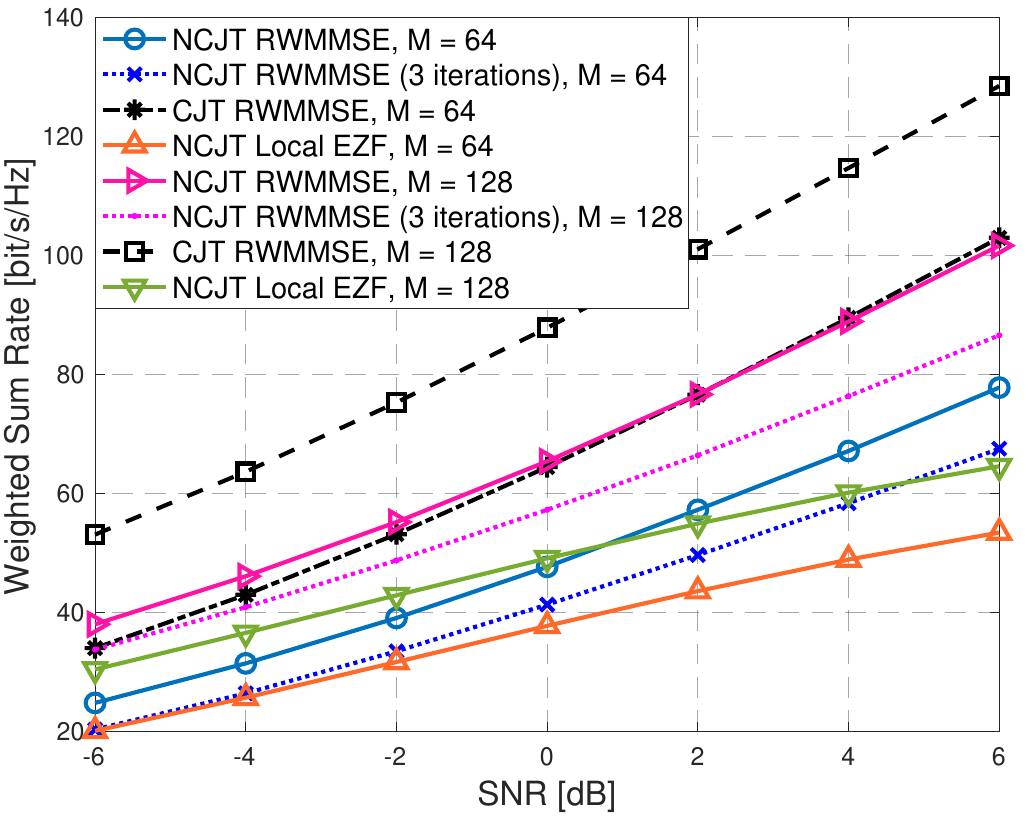}
\vspace{-10bp}
\caption{Comparison of CJT and NCJT with different numbers of transmit antennas ($M = 64,128$).}
\label{Comparison_CJT_NCJT}
\vspace{-5bp}
\end{figure}
Fig. \ref{Comparison_CJT_NCJT} exhibits the performance gaps between the CJT and the NCJT strategy under different numbers of transmit antennas. 
We can see that the CJT strategy provides approximately 30$\%$ more WSR than the NCJT strategy, since in CJT the APs cooperate as a virtue MIMO system and transmit the same signals to their serving UEs. 
However, CJT requires strict synchronization among the APs, which can be hard to achieve in practical communication systems. 
In contrast, NCJT avoids synchronization overhead such as pilots at an acceptable cost of the WSR performance.
In addition, our proposed RWMMSE algorithm achieves no less than $80\%$ WSR performance in 3 iterations,
showing great engineering prospect. 
%
%

\begin{figure}[!t]
\centering
\includegraphics[width=0.68\columnwidth]{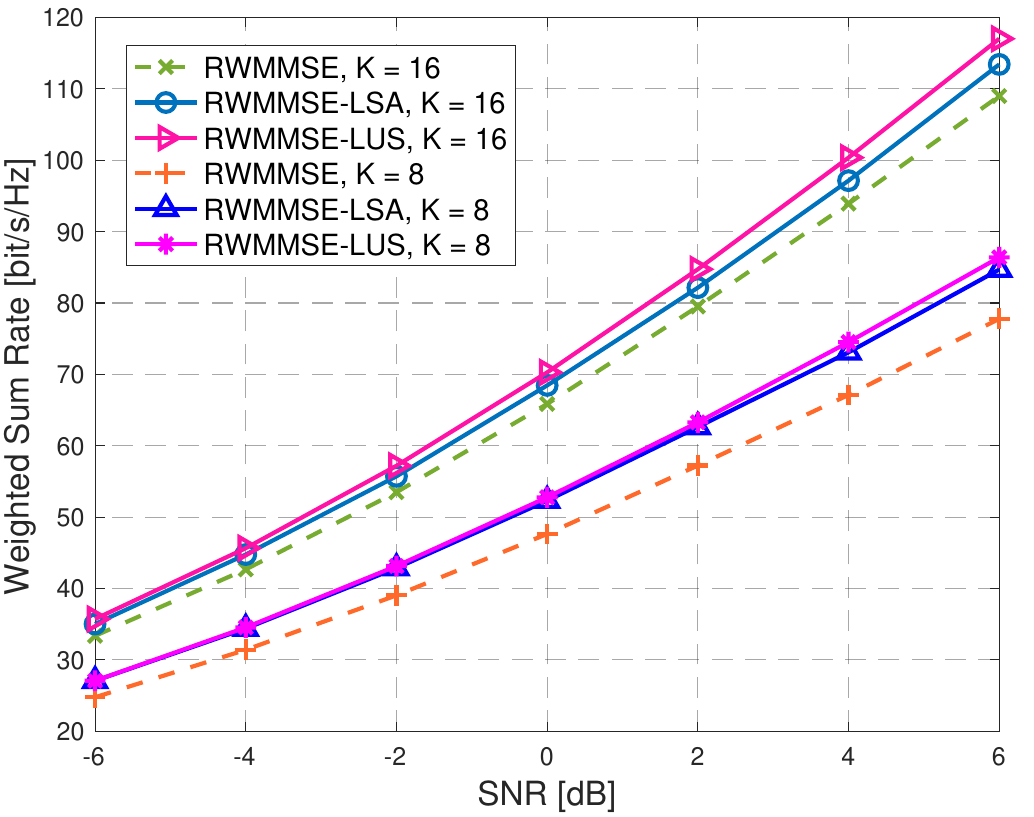}
\vspace{-10bp}
\caption{WSR Performance of the proposed RWMMSE-LSA and RWMMSE-LUS algorithms.}
\label{Performance_Lcon}
\vspace{-8pt}
\end{figure}
\vspace{-10pt}
\subsection{Stream Allocation Performance}
We consider the RWMMSE algorithm with an average number of streams, i.e., $D=2$, transmitted from the nearest APs $\mathcal{I}_k$ to UE $k$ as our benchmark.
In Fig.~\ref{Performance_Lcon}, we compare the WSR of the RMMMSE algorithm, the RWMMSE-LSA algorithm, and the RWMMSE-LUS algorithm, under different values of SNR and numbers of UEs. 
We observe that the RWMMSE-LUS algorithm achieves the highest WSR while the RWMMSE algorithm achieves the lowest WSR, showing the effectiveness of our proposed stream allocation algorithms.
This is because that the RWMMSE algorithm allocates APs $\mathcal{I}_k$ to transmit an average number of streams to each UE $k$, the  RWMMSE-LSA algorithm optimizes the number of data streams transmitted from APs $\mathcal{I}_k$ to maximize WSR, while the RWMMSE-LUS algorithm selects data streams among all the APs that contribute to the highest WSR.

\section{Conclusions}\label{sec_conclu}
In this work, we have investigated joint beamforming and stream allocation algorithms for WSR maximization in user-centric cell-free MIMO networks with NCJT.
We first propose a distributed low-interaction and low-complexity RWMMSE beamforming algorithm with closed-form updates in each iteration for the case of fixed streams.
Our proposed RWMMSE algorithm exhibits the lowest known complexity and requires much lower interaction across the networks, with no compromise to WSR performance. 
We further propose a zero-interaction Local EZF method based only on the local CSI to initialize the RWMMSE iteration.
Finally, by effectively decouple the beamforming and stream allocation variables, we develop a joint beamforming and linear stream allocation RWMMSE-LSA algorithm with linear stream allocation complexity for WSR maximization with varying streams.
Simulation results demonstrate the significant advantages of our proposed algorithms over the current best alternatives, in terms of both convergence time and WSR performance. 




\vspace{-5pt}
\begin{appendices}
\section{Proof of Theorem \ref{theorem_sum}} \label{Appendix0}
We use contradiction to prove \textbf{Theorem} \ref{theorem_sum}.
Since for any AP~$i$, the columns of $\bar{\bf{H}}_i^H$ are linearly independent, 
we define ${\tilde{\bf{H}}_{i,k,n}}$ consisting all columns of $\bar{\bf{H}}_i^H$ except the $n$th column ${\bf{h}}_{i,k,n}$, i.e.,
\setlength{\abovedisplayskip}{2pt}
\begin{equation}
	{\tilde{\bf{H}}_{i,k,n}} \triangleq \left[ {\bf{H}}_{i,1}^H,\dots,{\bf{H}}_{i,k-1}^H, \hat{{\bf{H}}}_{i,k}, {\bf{H}}_{i,k+1}^H, \dots,{\bf{H}}_{i,K}^H\right]
\end{equation}
\setlength{\belowdisplayskip}{2pt}
where 
$\hat{{\bf{H}}}_{i,k} \triangleq \left[ {\bf{h}}_{i,k,1},\dots,{\bf{h}}_{i,k,n-1},{\bf{h}}_{i,k,n+1}, \dots, {\bf{h}}_{i,k,N_k} \right]$, and the columns of ${\tilde{\bf{H}}_{i,k,n}}$ are linearly independent.

Assume that for AP $i$, the local optimal beamformer is $\{ {\bf{P}}_{i,k}^*, k \in \mathcal{U}_i \}$, and $\sum_{k\in \mathcal{U}_i}
\left\| {\mathbf{P}}_{i,k}^* \right\|_{\text{F}}^2 <  P_{\max,i }$. 
Then we introduce another set of beamformers $\{ \hat{\bf{P}}_{i,k}, k \in \mathcal{U}_i \}$, and the $d$th column of $\hat{\bf{P}}_{i,k}$ equals to that of ${\bf{P}}_{i,k}^*$, i.e., $\hat{\bf{p}}_{i,k,d} = {\bf{p}}_{i,k,d}^* $ for all $d = 1,\dots,D_{i,k}$ and $d \neq s$. The $s$th column of $\hat{\bf{P}}_{i,k}$ is defined as
\begin{equation}
	\hat{\bf{p}}_{i,k,s} \triangleq {\bf{p}}_{i,k,s}^* + \beta e^{j\theta}\triangle{\bf{p}}_{i,k,s}
\end{equation}
where $\beta >0$ is a scaling factor such that $\sum_{k\in \mathcal{U}_i} \| \hat{\mathbf{P}}_{i,k} \|_{\text{F}}^2 =  P_{\max,i }$, $\theta =  \angle\left({\bf{h}}_{i,k,n}^H, {\bf{p}}_{i,k,s}^* \right)$, and $\triangle{\bf{p}}_{i,k,s} = \prod_{\tilde{\bf{H}}_{i,k,n}}^{\bot}{\bf{h}}_{i,k,n}$.
In addition, as $\prod_{\tilde{\bf{H}}_{i,k,n}}^{\bot} \succ \mathbf{0}$, then we have ${\bf{h}}_{i,k,n}^H\triangle{\bf{p}}_{i,k,s} > 0$, and ${\bf{h}}_{i,l,n}^H\triangle{\bf{p}}_{i,k,s} = 0$ for any $l \in \mathcal{U}_{-k}$. 

Furthermore, the $n$th diagonal element of ${\bf{H}}_{i,l}\hat{\bf{P}}_{i,k}\hat{\bf{P}}_{i,k}^H{\bf{H}}_{i,l}^H$ is $| {\bf{h}}_{i,l,n}^H \sum_{d=1}^{D_{i,k}} \hat{\bf{p}}_{i,k,d} |^2$. And we have
\begin{equation}
		\left| {\bf{h}}_{i,l,n}^H \hat{\bf{p}}_{i,k,d} \right| 
		= \left| {\bf{h}}_{i,l,n}^H {\bf{p}}_{i,k,d}^* \right| , \text{if}~d\neq s ~\text{or}~l\neq k,
\end{equation}
and
\begin{equation}
\begin{aligned}
		\left| {\bf{h}}_{i,k,n}^H \hat{\bf{p}}_{i,k,s} \right| 
		& = \left| {\bf{h}}_{i,l,n}^H {\bf{p}}_{i,k,s}^*  + \beta e^{j\theta}{\bf{h}}_{i,l,n}^H\triangle{\bf{p}}_{i,k,s}\right| \\
		& = \left| {\bf{h}}_{i,l,n}^H {\bf{p}}_{i,k,s}^*\right|  + \beta {\bf{h}}_{i,l,n}^H\triangle{\bf{p}}_{i,k,s} \\
		& > \left| {\bf{h}}_{i,l,n}^H {\bf{p}}_{i,k,s}^*\right|.
		\end{aligned}
\end{equation}
Consequently, $R_k\left( {\hat{\mathbf{P}}_{i,k}} \right) > R_k\left( {\mathbf{P}_{i,k}^*} \right)$ and for $l \neq k$, $R_l\left( {\hat{\mathbf{P}}_{i,k}} \right) = R_l\left( {\mathbf{P}_{i,k}^*} \right)$.
 Plus, when $\beta = 0$, ${\hat{\mathbf{P}}}_{i,k}^* = {\mathbf{P}_{i,k}^*} $, as to $\beta\rightarrow 0$ and $\beta \neq 0$, ${\hat{\mathbf{P}}}_{i,k}^* \rightarrow {\mathbf{P}_{i,k}^*} $.
That is to say, $\left\{ {\hat{\mathbf{P}}_{i,k}} \right\}$ contributes more rate then $\left\{ {\mathbf{P}_{i,k}^*} \right\}$, which contradicts to the assumption that $\{ {\bf{P}}_{i,k}^*\}$ is the local optimal beamformer. Hence, to get the full WSR, $\sum_{k\in \mathcal{U}_i}
\left\| {\mathbf{P}}_{i,k}^* \right\|_{\text{F}}^2 =  P_{\max,i }$ always holds for any $i \in \mathcal{I}$. Then the proof is completed.
\hfill $\blacksquare$
%
%
%
%
%
%
\section{Proof of Proposition \ref{Pro_1}}\label{Appendix_proof_proposition}
By contradiction, suppose that $\mathbf{{P}}_{i,k}^*$ does not lie in the column space of $\bar{{\bf{H}}}_i^H$, then $\mathbf{P}_{i,k}^*$ can be expressed as
\begin{equation}
\mathbf{P}_{i,k}^*=\tilde{\mathbf{P}}_{i,k}+\hat{\mathbf{P}}_{i,k}, \forall k \in \mathcal{U}_i  
\end{equation}
where $\tilde{\mathbf{P}}_{i,k}$ and $\hat{\mathbf{P}}_{i,k}$ are located in the column space and null space of $\bar{{\bf{H}}}_i^H$, respectively.
Note that $\tilde{\mathbf{P}}_{i,k}$ and $\mathbf{P}_{i,k}^*$ have the same WSR due to ${\mathbf{H}}_{i,k}\hat{\mathbf{P}}_{i,k} = {\mathbf{0}}$. 
In other words, removing the part of the beamforming matrix ${{\bf{P}}_{i,k}^*}$ in the null space of ${\bar{\mathbf{H}}_i}^{H}$ has no effect on the objective function value of \eqref{WSR_max_MU_simplified}.
\hfill $\blacksquare$
%
%
%
%
%
%
%
\section{Proof of Theorem \ref{theom2}}\label{Appendix_proof_X}
We prove by contradiction. 
According to \textbf{Proposition}~\ref{Pro_1}, if the local optimum $\mathbf{{P}}_{i,k}^{*}$ does not lie in the column space of $\bar{{\bf{H}}}_i^H$, then the new beamformer obtained by removing its part in the null space has the same WSR as $\mathbf{{P}}_{i,k}^{*}$. This contradicts the conclusion of full power property in Theorem~\ref{theorem_sum}.
\hfill $\blacksquare$
%
%
%
%
%
%
%
%
%
%
%
\vspace{-5pt}

\section{Proof of Lemma \ref{Prop_cons_remove}}\label{Appendix_Prop_cons_remove}
Since the beamformer $\bar{{\bf{P}}}_{i,k}^{*(r)}$ in the $r$th iteration is updated given the stream indicator matrix ${{\bf{L}}}_{i,k}^{*{(r-1)}}$ in the $(r-1)$th iteration, the $\mathbf{\Upsilon}_{i,k}$th columns of $\bar{{\bf{P}}}_{i,k}^{*(r)}$ are zero with $\mathbf{\Upsilon}_{i,k}$ being the indexes of the zero diagonal elements of ${{\bf{L}}}_{i,k}^{*{(r-1)}}$.
Hence, we have $\sum_{k\in \mathcal{U}_i}\!\!\operatorname{Tr}\!\left( \bar{\mathbf{P}}_{i,k}^{*(r)}\mathbf{L}_{i,k}^{*(r)}(\bar{\mathbf{P}}_{i,k}^{*(r)})^{H}\right)\leq\sum_{k\in \mathcal{U}_i}\!\!\operatorname{Tr}\!\left( \bar{\mathbf{P}}_{i,k}^{*(r)}\mathbf{L}_{i,k}^{*(r-1)}(\bar{\mathbf{P}}_{i,k}^{*(r)})^{H}\right)\leq P_{\max,i}$.
%
%
%
%
%
%
\section{Proof of Proposition \ref{Prop_convergence}}\label{Appendix_Prop_convergence}
The convergence of the RWMMSE-LSA to the stationary points of problem \eqref{Stream_limited_Problem_c} is guaranteed by the classic convergence theory of the BCD method \cite{bertsekas1997nonlinear}.
Moreover, by rewriting the 0-1 integer constraints \eqref{stream_d2} as $l_{i,k}^{(m)}(l_{i,k}^{(m)}-1)=0,  \forall k, \forall i,  m = 1,\dots,N_k$, we derive the KKT condition of problem \eqref{Stream_limited_Problem}.
Upon comparing the two KKT systems of problem \eqref{Stream_limited_Problem_c} and \eqref{Stream_limited_Problem}, it can be observed that any limit point $\left( {\bf{U}}_k^*, {\bf{W}}_k^*,  {{\bar{\bf{P}}_{i,k}^*}},  {{{\bf{L}}_{i,k}^*}} \right)$ of the iterative sequence generated by the RWMMSE-LSA  adheres to the KKT condition of problem \eqref{Stream_limited_Problem}, by using the fact that the diagonal elements of ${\bf{L}}_{i,k}^*$ must be zero or one. 
Hence, $\left( {\bf{U}}_k^*, {\bf{W}}_k^*,  {{\bar{\bf{P}}_{i,k}^*}},  {{{\bf{L}}_{i,k}^*}} \right)$ is both the stationary point of problem \eqref{Stream_limited_Problem} and its equivalent original problem \eqref{WSR_problem}.
\hfill $\blacksquare$
\end{appendices}

\bibliographystyle{IEEEtran}
\bibliography{reference}



\end{document}